\documentclass[nofootinbib,amsmath,amssymb,aps,pra,10pt,floatfix,showkeys]{revtex4-2}%

\usepackage{dcolumn}%
\usepackage{bm}%
\usepackage{color}
\usepackage{lipsum}
\usepackage{multirow}
\usepackage{url}
\usepackage{hyperref}
\usepackage{makecell}
\usepackage{graphicx} 
\usepackage{float}

\newcommand{\Si}{\mathop{\mathrm{Si}}}
\newcommand{\Ci}{\mathop{\mathrm{Ci}}}
\newcommand{\SF}{\mathop{\mathrm{SF}}}
\newcommand{\CF}{\mathop{\mathrm{CF}}}

\begin{document}

\title{Transition of dimuonium through foil}

\author{Abdaljalel Alizzi}
\email{abdaljalel90@gmail.com}
\affiliation{Novosibirsk State University, Novosibirsk 630 090, Russia}
\affiliation{Al Furat University, Deir-ez-Zor, Syrian Arab Republic}

\author{Artem Uskov}
\email{a.uskov+articles@alumni.nsu.ru}
\affiliation{Budker Institute of Nuclear Physics, Novosibirsk 630 090, Russia}

\author{Z.~K.~Silagadze}
\email{Z.K.Silagadze@inp.nsk.su}
\affiliation{Novosibirsk State University, Novosibirsk 630 090, Russia}
\affiliation{Budker Institute of Nuclear Physics, Novosibirsk 630 090, Russia}

\begin{abstract}
This article presents a study of the passage of dimuonium through the foil of ordinary matter. First, we provide an overview of how dimuonium is planned to be produced for such a type of experiment and how it is expected to interact with the ordinary atoms --- predominantly electromagnetically via the screened Coulomb potential of the atomic nuclei. Then, we describe the transport equations that represent the evolution of dimuonium states during the passage and their solution methods. Finally, for three different foils (Beryllium, Aluminium and Lead), we present the results of this study. To estimate the impact of uncertainties in the potential of a target atom, we study 15 different approximations of the atomic potential and show that the corresponding atomic-potential-model-dependent error in the yields of the low lying states of dimuonium is quite small within the framework of the applied Born approximation.
The convergence of the results after truncation of the infinite system of transport equations to the finite number of quantum states of dimuonium is also studied, and good convergence for the yields of low-lying states is demonstrated.
\end{abstract}

\maketitle

\section{introduction}
Dimonium (true muonium) is a hydrogen-like exotic elementary atom consisting of muon and antimuon. It was discussed long ago \cite{Baier, Budini, Hughes, Malenfant:1987tm}, but has not yet been observed experimentally. Due to the larger mass of the muon, dimuonium can be more sensitive to effects of new physics beyond the Standard Model than its relatives, positronium or muonium. The sensitivity to new physics is further enhanced in the case of ditauonium \cite{d’Enterria2022}. However, the rapid weak decay of $\tau$, unfortunately, makes it difficult to observe and study true tauonium \cite{Brodsky:2009gx}.

Being a purely leptonic system, the properties of dimuonium are largely determined by quantum electrodynamics (QED), with hadronic and electroweak contributions being under control \cite{Jentschura,Lamm:2016vtf,Lamm:2015lia}. This circumstance makes dimuonium a testing ground for QED bound-state calculations \cite{karshenboim2005}. Interestingly, a recent measurement of the $n=2$ fine structure of positronium revealed an unexpected discrepancy of more than 4 standard deviations with the QED prediction for the interval $2^3S_1\to2^3P_0$ \cite{Gurung:2020hms}. The reason for the discrepancy has not yet been identified \cite{Adkins2022}. A recent measurement of the $2^3S_1\to2^3P_2$ positronium interval is consistent  with the theory \cite{sheldon2023} indicating that the $2^3S_1\to2^3P_0$ discrepancy is most likely an experimental artifact, but the story is not over yet.

There have been many interesting proposals for the production of dimuonium (see, for example, \cite{Malenfant:1987tm,Brodsky:2009gx,Bilenky:1969zd,Banburski:2012tk,Holvik:1986ty,Nemenov:1972ph,Ginzburg:1998df,Arteaga-Romero:2000mwd,Lamm:2017ioi} and references therein), but none of them have been implemented in practice. Perhaps the recent Novosibirsk project of the so-called $\mu\mu$-tron machine \cite{Bogomyagkov:2017uul} and its twin DIMUS at Fermilab \cite{Fox:2021mdn} have a better chance of implementation. Only $S$-state dimuoniums will be born at the $\mu\mu$-tron type machine. Other quantum states of dimuonium can be generated when dimuonium passes through a thin foil. The purpose of this work is to study this process.

When crossing the foil, dimuonium interacts with target atoms electromagnetically via the screened Coulomb potential $U(r)$. As a result, dimuonium either ionizes (dissociates into a free muon-antimuon pair) or passes from one quantum state to another state. Transitions between ortho and para states are significantly suppressed \cite{Denisenko:1987gr}. Therefore, and because only ortho states are produced at the $\mu\mu$-tron type machine through the one-photon annihilation of an electron and positron, when considering the evolution of quantum states of dimuonium in foil, we will limit ourselves to ortho states.

The corresponding discrete-discrete $(n,l,m)\to(n^{\prime},l^{\prime},m^{\prime})$ transition cross section, and the total $(n,l,m)\to X$ cross section have the form \cite{Mrowczynski:1985qt,Mrowczynski:1987gq,Denisenko:1987gr,Alizzi2021AlternativeIO} (for an alternative approach, see \cite{Halabuka_1999}):
\begin{equation}
d\sigma_{nlm}^{n^{\prime}l^{\prime}m^{\prime}}=\frac{e^{2}\left(1-(-1)^{l-l^{\prime}}\right)}{\pi V^{2}}\,\left|\tilde{U}(\vec{q})\right|^{2}\left|F_{nlm}^{n^{\prime}l^{\prime}m^{\prime}}\left(\frac{\vec{q}}{2}\right)\right|^{2}qdq,
\label{eq1}
\end{equation}
and
\begin{equation}
d\sigma_{nlm}^{tot}=\frac{e^{2}}{\pi V^{2}}\,\left|\tilde{U}(\vec{q})\right|^{2}\left[1-F_{nlm}^{nlm}(\vec{q}\,)\right]qdq.
\label{eq2}
\end{equation}
Here the atomic form factor $F_{nlm}^{n^{\prime}l^{\prime}m^{\prime}}$
is the Fourier transform of $\phi_{f}^{*}(\vec{r})\phi_{i}(\vec{r})$
with respect to the transferred momentum $q$, with $\phi_i(\vec{r})$ and $\phi_f(\vec{r})$ being the initial and final state Coulomb wave functions of the relative motion of muon and antimuon in the dimuonium atom. The Fourier image $\tilde{U}(q)$ of the atomic potential $U(r)$ in the case of a spherically symmetric atom can be calculated as follows \cite{Zhabitsky2014HowUI}:
\begin{equation}
\tilde{U}\left(q\right)=4\pi\int_{0}^{\infty}U\left(r\right)\frac{\sin{(qr)}}{qr}r^{2}dr=\frac{4\pi Ze}{q}\int_{0}^{\infty}\varphi\left(r\right)\sin{(qr)}dr. 
\label{eq3}
\end{equation}
The atomic screening function $0\le\varphi\left(r\right)\le1$ is defined by the relation
\begin{equation}
U\left(r\right)=Ze\frac{\varphi\left(r\right)}{r},
\label{eq4}
\end{equation}
and shows how atomic electrons screen the nuclear potential: it is zero for neutral atoms and one for the pure Coulomb potential of a bare nucleus. In other words, it is defined as the ratio between the electrostatic potential $U\left(r\right)$, experienced by an infinitesimal point charge at a distance $r$ from the nucleus (assuming the spherical symmetry), and the electrostatic potential of the bare nucleus \cite{Salvat1987AnalyticalDS}.

The atomic form factor $F_{nlm}^{n^{\prime}l^{\prime}m^{\prime}}$ can be expressed as follows \cite{Alizzi2021AlternativeIO}
\begin{equation}
F_{n_{1}l_{1}m_{1}}^{n_{2}l_{2}m_{2}}=N\sum\limits _{l=|l_{1}-l_{2}|}^{l_{1}+l_{2}}A_{l}\,I_{l}=N\sum\limits _{s=0}^{\mathrm{min}(l_{1},\,l_{2})}A_{l_1+l_2-2s}\,I_{l_1+l_2-2s},
\label{eq5}
\end{equation}
where 
\begin{equation}
N=\frac{(2a)^{l_{1}+1}\,(2b)^{l_{2}+1}}{n_{1}+n_{2}}\sqrt{(2l_{1}+1)(2l_{2}+1)\,\frac{(n_{1}-l_{1}-1)!\,(n_{2}-l_{2}-1)!}{(n_{1}+l_{1})!\,(n_{2}+l_{2})!}}.
\label{eq6}
\end{equation}
and ($Y_{lm}(\Omega_{q})$ are ordinary spherical functions, and Wigner $3j$-symbols arise after angular integration)
\begin{equation}
A_{l}=i^{l}(-1)^{m_{2}+m}\sqrt{4\pi(2l+1)}\left(\begin{array}{ccc}
l_{1} & l_{2} & l\\
0 & 0 & 0
\end{array}\right)\left(\begin{array}{ccc}
l_{1} & l_{2} & l\\
m_{1} & -m_{2} & -m
\end{array}\right)Y_{lm}(\Omega_{q}).
\label{eq7}
\end{equation}
The radial integral $I_{l}$  involves the spherical Bessel function $j$ and the product of two associated Laguerre polynomials $L$:
\begin{equation}
I_{l}=\int\limits _{0}^{\infty}x^{l_{1}+l_{2}+2}\,e^{-x}\,j_{l}(\sigma x)\,L_{n_{1}-l_{1}-1}^{2l_{1}+1}(2ax)\,L_{n_{2}-l_{2}-1}^{2l_{2}+1}(2bx)\,dx.
\label{eq8}
\end{equation}
The following notations were used in above expressions:
\begin{equation}
a=\frac{n_{2}}{n_{1}+n_{2}},\;\;b=\frac{n_{1}}{n_{1}+n_{2}},\;\;\sigma=\frac{n_{1}n_{2}}{n_{1}+n_{2}}\,q,\;\;x=\frac{r}{ab(n_{1}+n_{2})},\;\;s=\frac{1}{2}(l_1+l_2-l).\label{eq9}
\end{equation}
Three different methods for calculating the radial integral $I_{l}$ were described in \cite{Alizzi2021AlternativeIO}. A method by D.~P.~Dewangan \cite{Dewangan2012AsymptoticMF} is based on applying a variant of the Cauchy product formula of two associated Laguerre polynomials and then evaluating
some integral which can be found in the classical table of integrals by Gradshteyn and Ryzhik \cite{Gradshteyn1966TableOI}. As a result, this method evaluates the atomic form factor as a fourfold finite series of rational functions of the transferred momentum.

Another method by L. Afanasyev and A. Tarasov \cite{Afanasyev_1993,Afanasev1996BreakupOR}, which has increased computational efficiency, but is more complex than the Devangan method, is based on a Clebsch-Gordan type linearization relation for the product of two associated Laguerre polynomials with Jacobi polynomials as expansion coefficients. This method was implemented as a FORTRAN program in 2003, see \cite{Rios2003AnIO}.

An alternative method for calculating the radial integral was developed in \cite{Alizzi2021AlternativeIO}. It is  based on the mathematical result of \cite{Alassar2008ANI} on the integral involving the product of Bessel functions and associated Laguerre polynomials. The result is expressed in terms of the Gaussian hypergeometric function, which itself can be expressed in terms of Jacobi polynomials, and the latter can be calculated using a three-term recurrence function. This new method complements both Dewangan's method and Afanasyev and Tarasov's method in the sense that it combines the simplicity and directness of the former method with the computational efficiency of the latter. 

Our numerical results were obtained using the latter method. However, to reduce the likelihood of errors in the computer code, all three methods were implemented in a FORTRAN program with the Python interface \cite{Uskov2022MuMuPyAD} and were shown to give the same results for dimuonium cross sections.

It is convenient to choose the quantization axis parallel to the direction of the dimuonium beam, since if the initial momentum of a dimuonium atom is much larger than the transferred momentum during a collision, then the quantization axis remains practically unchanged in subsequent collisions \cite{Rios2003AnIO}. To obtain results that do not depend on the choice of the quantization axis, it is necessary to sum the transitions over magnetic quantum numbers.

Throughout this article, as in \cite{Alizzi2021AlternativeIO,Uskov2022MuMuPyAD}, we use mixed natural-atomic units in which $c=\hbar=1$ and the unit of mass is $\frac{1}{2}m_{\mu}$ (reduced mass of a dimuonium atom). In these units, the elementary electric charge is $e=\sqrt{\alpha}$, where $\alpha$ is the fine-structure constant, and the radius of the first Bohr orbit in dimuonium is $a_B=1/\alpha$. All lengths are expressed in terms of $a_B$ and to convert cross sections to ordinary units, it is useful to remember that $a_{B}=2\,\frac{m_{e}}{m_{ \mu}}\, a_{0}\approx 512~\mathrm{fm} $, where $a_{0}\approx5.29\times10^{-11}~\mathrm{m}$ is the usual Bohr radius.

\section{Transport equations}
In this section we describe the physics of the evolution of dimuonium states when passing through a foil of ordinary matter. Dimuonium at $\mu\mu$-tron type machines is produced in electron-positron annihilation and to a good approximation its production cross-section is proportional to the squared modulus of the dimuonium wave function at the origin. Thus, in the non-relativistic approximation, only $S$-states of  orthodimuonium are born, and the prevalence of the resulting $nS$-states is proportional to $1/n^3$, since
\begin{equation}
\mid\Psi_{n00}(0)\mid^{2}=\frac{1}{\pi a_B^3 n^{3}}=\frac{\alpha^3 m_\mu^3}{8\pi n^3}.
\label{eq10}
\end{equation}
But immediately after dimuonium $nS$ states have been created and flown forward from the origin, they undergo a natural decay process $(\mu^+\mu^-)\to e^+e^-$ and their number after traveling a distance $l$ is determined by the formula:
\begin{equation}
N_{n00}(l)=\frac{1}{n^{3}}e^{-l/(V\gamma\tau)}N_{0}=\frac{1}{n^{3}}e^{-l/l_n}N_{0},
\label{eq11}
\end{equation}
where $l_{n}=V\gamma\tau$ is the characteristic natural-decay length of the $nS$ state, and $N_{0}$ is the number of initially produced dimuonium $1S$ states at the electron-positron interaction point. If the crossing angle of electron and positron beams is $\theta$, the created dimuonium moves with the velocity $\beta=V/c=\beta_e\sin{\theta}$ \cite{Uskov2022MuMuPyAD}. In the $\mu\mu$-tron project the designed beam energy is 408~MeV and the crossing angle $\theta=75^\circ$. Then $\beta_e\approx 1$, $\beta\gamma=\tan{\theta}\approx 3.73$ and since QED predicts $c\tau=0.543~\mathrm{mm}$ \cite{Brodsky:2009gx,Bilenky:1969zd} for the orthodimuonium $1S$ state, we get $l_n\approx n^3\, 2.03~\mathrm{mm}$ (thanks to (\ref{eq10}), the $nS$ state lives $n^3$ times longer than the $1S$ state). If the foil is placed at a distance $d$ from the interaction point, the initial conditions for the evolution of quantum states of dimuonium inside the foil will be 
\begin{equation}
N_{nlm}(0)=\left \{ \begin{array}{cc} \frac{N_0}{n^3}e^{-\frac{d}{2.03n^3}}, & \mathrm{if}\;\; (n,l,m)=(n,0,0), \\
0, & \mathrm{if}\;\; (n,l,m)\ne (n,0,0), \end{array} \right .
\label{eq12}
\end{equation}
where now $N_{nlm}(0)$ is the number of $(n,l,m)$-states  when the dimuonium beam enters the foil (the zero of its argument corresponds to the distance traveled by the dimuonium in the foil).

The transport of an elementary atom in matter is usually considered using the classical probabilistic picture and the corresponding kinetic equations that neglect quantum interference between degenerate states with the same energy \cite{Afanasev1996BreakupOR,Afanasyev:2004pq}. In our case, the transport equations have the form
\begin{equation}
    \frac{dN_\alpha}{dl}=-n_Z\sigma(\alpha\to X)N_\alpha+\sum\limits_{\beta\ne\alpha}n_Z\sigma(\beta \to \alpha)N_\beta,
    \label{eq13}
\end{equation}
where $\alpha=(n,l,m)$ denotes the quantum state of dimuonium and $n_Z$ is the number of target atoms per unit volume.

The transport equations (\ref{eq13}) describe the evolution of dimuonium quantum states (their occupation numbers $N_\alpha$) as a function of the distance $l$ traveled through the foil. The abundance of any particular state through the foil is controlled by a negative term describing its dissociation or excitation into other states, and an infinite number of positive terms describing the formation of that state through all possible transitions from other states.

It is convenient to express $l$, the thickness of the target foil through the dimensionless quantity $z=l/l_{1S}$  \cite{Banburski:2012tk}, where  $l_{1S}$ is the characteristic in-target-decay-length of the $1S$ state, for which we have:
\begin{equation}
l_{1S}=\frac{1}{n_Z\sigma(1S\rightarrow X)}.
\label{eq14}
\end{equation}
Then the transport equations (\ref{eq13}) will take the form:
\begin{equation}
    \frac{dY_\alpha}{dz}=-\frac{\sigma(\alpha\to X)}{\sigma(1S\to X)}Y_\alpha+\sum\limits_{\beta\ne\alpha}\frac{\sigma(\beta \to \alpha)}{\sigma(1S\to X)}Y_\beta,
    \label{eq15}
\end{equation}
where $Y_\alpha=N_\alpha/N_{0}$ expresses the yield of the state $\alpha=(n,l,m)$ as a fractional ratio of $N_\alpha$ to $N_{0}$.

Since the $Z$ dependence is significantly leveled out in cross-section ratios, the transport equations in such a dimensionless form are less sensitive to the type of target material.

Before moving on, there are several important things to mention.
\begin{itemize}
\item The natural decay process of orthodimuonium $(\mu^+\mu^-)\to e^+e^-$, of course, continues when dimuonium passes through the foil, but the corresponding terms are not included in the transport equations. Such terms appear predominantly only for $nS$ states and to include them, according to (\ref{eq11}), we have to make a replacement
\begin{equation}
 \frac{\sigma(\alpha\to X)}{\sigma(1S\to X)} \to  \frac{\sigma(\alpha\to X)}{\sigma(1S\to X)}+\frac{1}{n_Z\sigma(1S\to X)l_n}=
 \frac{\sigma(\alpha\to X)}{\sigma(1S\to X)}+\frac{l_{1S}}{l_1 n^3}
 \label{eq16}
\end{equation}
in the first term on the right side of the equations (\ref{eq15}). Since $n_Z=\rho N_A/A$, where $\rho$ is the density, $A$ is the molar mass and $N_A$ is Avogadro’s number, we find accordingly that $l_{1S}=4.4~\mu\mathrm {m}$ for lead ($A=207.2~\mathrm{g}/\mathrm{mol}$, $\rho=11.35~\mathrm{g}/\mathrm{cm}^3$, $\sigma( 1S\to X )\approx 6.9\cdot 10^{-20}~\mathrm{cm}^2$), $l_{1S}=79.2~\mu\mathrm{m}$ for aluminum ($A=26.982~\mathrm {g}/\mathrm{mol}$, $\rho=2.7~\mathrm{g}/\mathrm{cm}^3$, $\sigma(1S\to X)\approx 2.09\cdot 10^{-21}~\mathrm{cm}^2$), and  $l_{1S}=0.37~\mathrm{mm}$ for beryllium ($A=9.01218~\mathrm{g}/\mathrm{mol}$, $ \rho=1.85 ~\mathrm{g}/\mathrm{cm}^3$, $\sigma(1S\to X)\approx 2.2\cdot 10^{-22}~\mathrm{cm}^2 $). Dividing by $l_1\approx 2.03~\mathrm{mm}$, we obtain the following values for the corresponding correction terms (for the $1S$ state): $2.17\cdot 10^{-3}$, $3.9\cdot 10^{-2}$ and $0.18$, respectively. Since $\sigma(nS\to X)/\sigma(1S\to X)\ge 1$, we see that such a correction has a noticeable significance only for materials with low $Z$ and only for $1S$ and (probably) $2S$ quantum states.

\item Transport equations (\ref{eq15}) are an infinite system of linear differential equations. In practice we have to truncate it to a finite number of equations involved. As a truncation parameter, we choose $n_{max}$ -- the maximum principal quantum number for the states $\alpha=(n,l,m)$ included in the transport. Then the total number of active $\alpha=(n,l,m)$ states included in the truncated transport equations, and hence the dimension of the truncated system (\ref{eq15}), will be equal to
\begin{equation}
    N=\sum\limits_{n=1}^{n_{max}}n^2=\frac{n_{max}(n_{max}+1)(2n_{max}+1)}{6}.
    \label{eq17}
\end{equation}
In fact, not all transitions between these states are allowed. In addition to the selection rule $(-1)^{l-l^\prime}\ne 1$, which follows from (\ref{eq1}), we have a less obvious selection rule related to the conservation of the so-called $Z$-parity $P_Z=(-1)^{l-m}$, when the quantization axis is along the dimuonium beam direction \cite{Christova,Afanasev1996BreakupOR,Alizzi2021AlternativeIO}. We did not use this opportunity to reduce the dimension of the truncated transport equations, since the numerical solution up to $n_{max}=10$ (a total of 385 $\alpha=(n,l,m)$ states) did not cause any difficulties.

To detect and correct possible errors in computer codes, we implemented two different methods for numerically solving truncated transport equations. In FORTRAN code, we use the DDEQMR (D202) program from the CERN Program Library (CERNLIB) to solve a system of first-order differential equations using the Runge-Kutta-Merson method. Alternatively, standard Python tools were used to find the eigenvalues (assumed to be non-degenerate) and eigenvectors of the corresponding coefficient matrix, and the solution to the system of linear differential equations was expressed in terms of the matrix exponent (in general, calculating the matrix exponent numerically can be quite a subtle task \cite{Moler}, but we did not encounter problems with numerical instability).

\item According to the principle of uncertainty, the quantum state of an elementary atom with $E_n$ energy requires time $\tau \sim \hbar/|E_n|$ for formation. Since this formation time is very small (for $n=1$, $\tau\sim 0.97\cdot 10^{-16}~\mathrm{s}$ for positronium and $\tau\sim 0.47\cdot 10^{-18}~\mathrm{s}$ for dimuonium), it is usually not taken into account. However, for ultrarelativistic elementary atoms with very high $\gamma$-factors, it may happen that the time of crossing the foil will be much less than the formation time  $\gamma\tau$ in the laboratory system. In this case, the phenomenon of so-called superpenetrability arises \cite{Nemenov:1981kz,Nemenov:1989nx} and the above transport equations can no longer be applied. In our case, this interesting effect is irrelevant, since dimuonium is assumed to be only mildly relativistic.

\item Although the form of the transport equations (\ref{eq15}) resembles the Pauli Master equation from quantum statistical mechanics \cite{Zubarev}, they are classically probabilistic in nature, since they ignore any quantum interference between different states of dimuonium as it passes through matter. These interference effects play a decisive role in the occurrence of superpenetration phenomena at ultrarelativistic energies \cite{Lyuboshits:1981ky}, but are usually ignored if $\gamma$-factors are not very high. However, it has been suggested that for hydrogen-like elementary atoms, some interference effects may occur even at low energies due to accidental degeneracy of the energy levels of hydrogen-like atoms, which increases the number of states among which interference can be significant \cite{Voskresenskaya_2003}. Accordingly, to describe the passage of an elementary atom through matter, a more complex approach was proposed, based on the description of quantum systems using a density matrix \cite{Voskresenskaya_2003}. In the case of pionium, a $\pi^+\pi^-$ hydrogen-like atom, the evolution of its states when passing through matter was studied within the framework of the density matrix formalism, but no significant differences from the classical probabilistic picture were found within the specific conditions of the Dimeson Relativistic Atom Complex (DIRAC) experiment at CERN \cite{Afanasyev:2004pq}.

\item In addition to the static atomic potential $U(r)$, in principle it is also necessary to take into account polarization potentials (van der Waals forces) \cite{Krachkov:2023jcj}. If ${\bf{r_1}}$ and ${\bf{r_2}}$ are the radius vectors of the muon and antimuon, then the effects of atomic polarization are effectively taken into account by replacing $e[U(r_2)-U(r_1)]$ in the interaction Hamiltonian by
\begin{equation}
e[U(r_2)-U(r_1)]+U_{pol_1}(r_1)+U_{pol_1}(r_2)+U_{pol_2}({\bf{r_1}},{\bf{r_2}}).
\label{eq18} 
\end{equation}
Typically, a semi-empirical method is used to determine polarization potentials. For example, Biermann one-body polarization potential has the form \cite{Chisholm_1964}
\begin{equation}
    U_{pol_1}(r)=-\frac{\alpha_d}{2r^4}\left (1-e^{-r^8/r_0^8}\right ),
    \label{eq19}
\end{equation}
where $\alpha_d$ is the static dipole polarizability of the atom, and $r_0$ is the empirical cutoff parameter. The two-body polarization potential is necessary because the polarization of an atom by one body affects the potential experienced by the other. It was first introduced by Chisholm and  {\"{O}}pik in the form \cite{Chisholm_1964}
\begin{equation}
U_{pol_2}({\bf{r_1}},{\bf{r_2}}) =\frac{\alpha_d}{r_1^3r_2^3}({\bf{r_1}}\cdot{\bf{r_2}})\left (1-e^{-r_1^8/r_0^8}\right )\left (1-e^{-r_2^8/r_0^8}\right ).
\label{eq20}
\end{equation}
More refined polarization potentials can be found in \cite{Stewart}. We expect the polarization potential corrections to be small in our case since dimuonium is an electrically neutral system.

\item The total and transition cross sections used in this work are calculated in the Born approximation, as in most similar studies, and in confirmation of the old wisdom: ``As the bee seeks the honey, so do theoretical physicists seek cross sections not yet computed in Born approximation, because Born approximation usually is comparatively easy to calculate, whereas anything better is usually much harder" \cite{Gerjuoy_1964}. Rough criterion for the validity of the Born approximation is that the velocity of the incident projectile must be large compared to the velocities of bound electrons in the target atom or, which is very approximately the same, that the kinetic energy of the incident projectile must be large compared to the interaction energy \cite{Gerjuoy_1964}.  At first sight, in our case these criteria are satisfied. However, if the goal is to calculate cross sections with an accuracy of 1\%, it is necessary to take into account relativistic and multiphoton exchange corrections \cite{Santamarina_2003,Afanasyev:2004pq}.

\item This work does not take into account the possibility of excitation of a target atom during a collision. In the case of excitation of a target atom, the contributions of bound electrons must be summed incoherently, since the excitation affects the electrons individually. Therefore, this incoherent part of the cross section being proportional to the number of bound electrons is expected to be roughly $Z$-times smaller than the coherent part which is proportional to $Z^2$ (the charge of the nucleus squared). However, in general, this simple scaling argument is not accurate enough, and at the 1\% accuracy level it is important to account for the incoherent contribution even in the case of moderately heavy elements such as Ti or Ni \cite{Heim_2000}.

\end{itemize}

\section{Dependency of Cross Sections on the approximation of the atomic potential}

In order to investigate the dependence of cross sections on the approximation of the atomic potential, we have considered various approximations based on different models of atomic structure. In the context of the DIRAC experiment, a similar study was performed in \cite{Zhabitsky2014HowUI}. 

A slightly modified version of our FORTRAN program and MuMuPy calculator \cite{Uskov2022MuMuPyAD} that includes Fourier images of the new atomic potentials was used for this goal. The tested approximations can be divided into three groups.

The first group includes analytical approximations of the numerical solution of the Thomas-Fermi equation. It is known that the Thomas-Fermi model is valid in the average range of distances from the atomic nucleus and for a sufficiently large number of atomic electrons to satisfy the requirements of the statistical nature of the model \cite{Jackiw}. 

In the Thomas-Fermi model screening functions, the distance from the nucleus is expressed in dimensionless units as follows:
\begin{equation}
x=\frac{r}{b},\;\;b=b_ca_0,\;\;b_c=\left(\frac{9\pi^{2}}{128 Z}\right)^{1/3},
\label{eq21}   
\end{equation}
where $b$ is the Thomas-Fermi screening length.

The Moli\'{e}re approximation is a paradigmatic example of an analytical approximation of the Thomas-Fermi screening function, often used in various applications \cite{Moliere}:
\begin{equation}
\varphi(x)=\sum\limits_{i=1}^3 A_ie^{-\beta_i x},
\label{eq22}   
\end{equation}
where
\begin{equation}
A_1=0.35,\;A_2=0.55,\;A_3=1-A_1-A_2,\;\;
\beta_1=0.3,\;\beta_2=4\beta_1,\;\beta_3=20\beta_1.
\label{eq23}   
\end{equation}

The Fourier image of the Thomas-Fermi-Moli\'{e}re atomic potential in our units reads:
\begin{equation}
\tilde{U}_{MOL}(\tilde{q})=4\pi Z\sqrt{\alpha} \, a_B^2\,\sum\limits_{i=1}^3 \frac{A_i}{\tilde{q}^2+\tilde{\beta}_i^2},\;\; \tilde{\beta}_i=
2\,\frac{m_e}{m_\mu}\,\frac{\beta_i}{b_c},\;\;\tilde{q}=q\,a_B.
\label{eq24}   
\end{equation}
The Rozental approximation \cite{Rozental_1936,A_Jablonski_1981} is the same as (\ref{eq22}), but with a different set of parameters:
\begin{equation}
A_1=0.255,\;A_2=0.581,\;A_3=1-A_1-A_2,\;\;
\beta_1=0.246,\;\beta_2=0.947,\;\beta_3=4.356.
\label{eq25}   
\end{equation}
The corresponding Fourier transform $\tilde{U}_{ROZ}$ is similar to (\ref{eq24}) with obvious substitutions.

The two-parameter Csavinszky  parameterization has the form \cite{Csavinszky,Porta2016ComparisonOV}
\begin{equation}
\varphi_{CS}(x)=\left (\sum\limits _{i=1}^{2}a_{i}e^{-\beta_{i}x}\right )^{2},
\label{eq26}
\end{equation}
with
\begin{equation}
a_1=0.7218337,\;\;a_2=1-a_1,\;\; \beta_1=0.1782559, \;\;\beta_2=1.759339.
\label{eq27}
\end{equation}
Later, Kesarwani and Varshni proposed a similar three-parameter parameterization \cite{Kesarwani,Porta2016ComparisonOV}:
\begin{equation}
\varphi_{KV}(x)=\left (\sum\limits _{i=1}^{3}a_{i}e^{-\beta_{i}x}\right )^{2},
\label{eq28}
\end{equation}
with
\begin{eqnarray} &&
a_1=0.52495, \; a_2=0.43505, \; a_3=0.04, \nonumber \\ &&
\beta_1=0.12062,  \; \beta_2=0.84795, \; \beta_3=6.7469.
\label{eq29}
\end{eqnarray}
The Fourier images $\tilde{U}_{CS}$ and $\tilde{U}_{KV}$ are trivial generalizations of (\ref{eq24}) and can be implemented without difficulty.

Implementing the other two approximations of the Thomas-Fermi screening function requires more care because they are not exponential like the Moli\'{e}re approximation. 

Roberts proposed the following one-parameter parameterization \cite{Roberts,Porta2016ComparisonOV}
\begin{equation}
\varphi_{ROB}\left(x\right)=(1+a\sqrt{x})e^{-a\sqrt{x}},\;\;a=1.905.
\label{eq30}
\end{equation}
The Fourier image of the Roberts potential is (see appendix \ref{AppA})
\begin{equation}
\tilde{U}_{ROB}(\tilde{q})=\frac{4\pi Z\sqrt{\alpha}a_B^2}{\tilde{q}^{2}} \left\{ 1-\frac{\eta^{3}}{2}\left [\sin\left (\frac{\eta^{2}}{4}\right )\left (\sqrt{\frac{\pi}{8}}-\SF\left (\frac{\eta}{2}\right )\right )+\cos\left (\frac{\eta^2}{4}\right )\left (\sqrt{\frac{\pi}{8}}-\CF\left (\frac{\eta}{2}\right )\right )\right ]\right \},
\label{eq31}
\end{equation}
where
\begin{equation}
\eta=\frac{\bar{\eta}}{\sqrt{\tilde{q}}},\;\;\bar{\eta}=\sqrt{\frac{2m_e}{m_\mu b_c}}\,a,
\label{eq32}
\end{equation}
and $\SF(x),\CF(x)$ are Fresnel integrals, and we use the following definitions:
\begin{equation}
\SF(x)=\int\limits_0^x \sin{(t^2)}\,dt,\;\;\CF(x)=\int\limits_0^x\cos{(t^2)}\,dt.
\label{eq33}  
\end{equation}
For small $\tilde{q}=a_B q$ (dimensionless transferred momentum in our units), one can use the asymptotic expansion of the  Fresnel auxiliary function 
$g(z)$ (formula 7.3.28 in \cite{Abramowitz}. Be aware of a different definition of Fresnel integrals in  \cite{Abramowitz}) and get
\begin{equation}
\tilde{U}_{ROB}(\tilde{q})\approx 4\pi Z\sqrt{\alpha} a_B^2\left  [\frac{60}{\bar{\eta}^4}-\frac{15120}{\bar{\eta}^8}\,\tilde{q}^2+
\frac{15120\times 572}{\bar{\eta}^{12}}\,\tilde{q}^4\right ], \;\;{\mathrm{if}}\;\;\tilde{q}\ll 1.
\label{eq34}
\end{equation}

Another one-parameter parameterization was proposed by Tietz \cite{Tietz1955AnIA}:
\begin{equation}
\varphi_{Tietz}\left(x\right)=\frac{a_T^{2}}{\left(x+a_T\right)^{2}},\;\; a_T=\left(\frac{256}{35\pi}\right)^{2/3}\approx 1.75663.
\label{eq35}
\end{equation}
The Fourier transform of the Tietz potential is relatively easy to find and has the form 
\begin{equation}
\tilde{U}_{Tietz}(\tilde{q})=\frac{4\pi Z\sqrt{\alpha}a_B^2}{\tilde{a}^{2}}\left \{\sin\left (\frac{\tilde{q}}{\tilde{a}}\right )\left [\frac{\pi}{2}-\Si\left (\frac{\tilde{q}}{\tilde{a}}\right )\right ]-\cos\left (\frac{\tilde{q}}{\tilde{a}}\right )\Ci\left (\frac{\tilde{q}}{\tilde{a}}\right )\right \},
\label{eq36}
\end{equation}
where
\begin{equation}
\tilde{a}=\frac{2m_e}{m_\mu b_c a_T},\;\;\Si(x)=\int\limits_0^x\frac{\sin{t}}{t}dt,\;\;\Ci(x)=-\int\limits_x^\infty\frac{\sin{t}}{t}dt.
\label{eq37}
\end{equation}
For large arguments we can use asymptotic expansions \cite{Abramowitz}
\begin{eqnarray} &&
\frac{\pi}{2}-\Si(x)=f(x)\cos{x}+g(x)\sin{x}, \;\;\Ci(x)=f(x)\sin{x}-g(x)\cos{x},\nonumber \\ &&
f(x)\approx \frac{1}{x}\left (1-\frac{2}{x^2}+\frac{24}{x^4}\right),\;\;
g(x)\approx \frac{1}{x^2}\left (1-\frac{6}{x^2}+\frac{120}{x^4}\right),\;\;\mathrm{if}\;\; x\gg 1.
\label{eq38}
\end{eqnarray}
Exemplary results of calculating cross sections and their dependence on the type of approximation of the Thomas-Fermi potential are presented in the table \ref{Tab1}.
\begin{table}[H]
\begin{center}
\scriptsize
\begin{tabular}{|c|c|c|c|c|c|c|c|}
\hline 
$Z$ & \thead{Cross\\ Sections $[\mathrm{cm}^{2}]$}  & Roberts & \thead{Kesarwani \\ \& Varshni} & Moli\'{e}re & Rozental  & Csavinzky&
 Tietz
\tabularnewline
\hline 
\hline 
\multirow{2}{*}{$4$} & $\sigma(100\rightarrow211)\cdot 10^{23}$  & $5.322$ & $5.343$ & $5.351$ & $5.355$  & $5.389$ & $5.417$ \\
\cline{2-8} 
& $\sigma(100\rightarrow X)\cdot 10^{22}$   & $2.189$ & $2.197$ & $2.200$ & $2.202$ & $2.214$ & $2.224$
\tabularnewline
\hline 
\hline
\multirow{2}{*}{$13$} & $\sigma(100\rightarrow211)\cdot 10^{22}$   & $4.987$ & $5.006$ & $5.013$ & $5.017$ & $5.052$ & $5.080$
\tabularnewline
\cline{2-8} 
& $\sigma(100\rightarrow X)\cdot 10^{21}$  & $2.082$ & $2.090$ & $2.093$ & $2.094$  & $2.107$ & $2.118$
\tabularnewline
\hline 
\hline
\multirow{2}{*}{$82$} & $\sigma(100\rightarrow211)\cdot 10^{20}$ & $1.597$ & $1.601$ & $1.603$ & $1.604$  & $1.616$ & $1.626$
\tabularnewline
\cline{2-8} 
& $\sigma(100\rightarrow X)\cdot 10^{20}$ & $6.868$ & $6.892$ & $6.901$ & $6.905$  & $6.953$ & $6.991$
\tabularnewline
\hline 
\end{tabular}
\end{center}
\caption{The transition cross section $\sigma(100\rightarrow211)$
and the total cross section $\sigma(100\rightarrow X)$ for different approximations to the Thomas-Fermi potential.}
\label{Tab1}
\end{table}

The Thomas-Fermi approximation does not take into account the electron-electron interaction due to the Pauli principle (electrons with parallel spins avoid approaching each other, which leads to the appearance of exchange energy). We can estimate the relative importance of this effect using simple dimensional arguments \cite{Kirzhnits_1975,March_1975,Englert_1988}. Since the kinetic energy is proportional to $1/m$, its density, expressed in terms of the electron density $n$, has the form $\epsilon_K\sim n^\nu/m=e^2 a_0 n^\nu$, where $\nu=5/$3 for dimensional reasons. On the other hand, exchange energy originates from the electron-electron interactions which is $\sim e^2/r$. Therefore, its density $\epsilon_e\sim e^2 n^\mu$, where $\mu=4/3$ again on dimensional grounds. Thus, The relative importance of the exchange energy is $\delta_e\sim e^2 n^{4/3}/[e^2  a_0 n^{5/3}]=1/(a_0n^{1/3})$. But $n\sim Z/b^3$, and finally we get $\delta_e\sim (b/a_0)Z^{-1/3}\sim Z^{-2/3}$. When $Z^{2/3}$ is not a large number, many applications of the Thomas-Fermi model require such corrections to be taken into account. Exchange corrections were included in the Thomas-Fermi model by Dirac \cite{Dirac_1930,Jackiw}, and our second group of tested potentials includes analytical approximations of the numerical solution of the Thomas-Fermi-Dirac model.

One remark is appropriate here. Dirac obtained his approximation by neglecting the fact that momenta do not commute with coordinates in order to reduce the description of the atom to a semi-classical form \cite{Dirac_1930}. Quantum commutators lead to appearance terms in the energy functional which depend on $\nabla U$ (which can be traded for $\nabla n$). The magnitude of the corresponding quantum corrections can also be estimated from dimensional considerations \cite{Kirzhnits_1975,Englert_1988}. The quantum corrections are small if the electron density $n$ does not change significantly over a range on the order of the de Broglie wavelength $\lambda\sim 1/p_F\sim n^{-1/3}$. Consequently, the small parameter associated with these corrections is $\lambda |\nabla n|/n\sim n^{-4/3}|\nabla n|$. But $\nabla n$ is a vector, and energy is a scalar. Therefore, energy is a function of $(\nabla n)^2$, and the parameter describing the smallness of quantum correlations is equal to $\delta_q\sim n^{-8/3}(\nabla n)^2$. Since $|\nabla n|\sim n/b$, $n\sim Z/b^3$, we get $\delta_q\sim n^{-2/3}/b^2\sim Z^{-2/3}$.

As we see, the quantum corrections are of the same order $Z^{-2/3}$ as exchange corrections. However, a careful calculation (first performed in \cite{Kompaneets_1956}) shows that the quantum corrections come out with a small numerical coefficient and, as a result, only the exchange correction is noticeable in the Thomas-Fermi potential, giving the lion's share of $\sim Z^ {-2/3}$ corrections \cite{Englert_1988}.

Bonham and Strand \cite{Bonham,Salvat2005ELSEPAD} use Moli\'{e}re-like parameterization (\ref{eq22}) for analytical approximations to the numerical solution of the Thomas-Fermi-Dirac equation. Unlike the Thomas-Fermi case, the Thomas-Fermi-Dirac screening function does not show universal scaling in $Z$, and $A_i,\beta_i$ approximation parameters depend on $Z$ in a more complicated way. Bonham and Strand use fourth-order polynomial in $\ln{Z}$ to fit this dependence.

However, the Bonham and Strand approximation $\varphi_{BS}(x)$ does not satisfy the boundary conditions of the Thomas-Fermi-Dirac equation, which assumes a finite atomic size $x_0$ (depending on $Z$).  Jablonski proposed a modification of the  Bonham-Strand screening function to adjust it to the correct boundary conditions \cite{Jablonski1992ApproximationOT}:
\begin{equation}
\varphi_{TFDJ}(x)=\left\{ \begin{array}{c} \varphi_{BS}(x)+A(x/x_0)+B(x/x_0)^2,\;\;\mathrm{if}\;\;x\le x_0, \\
0, \;\;\mathrm{if}\;\;x> x_0. \end{array}\right .
\label{eq39} 
\end{equation}
The parameters $A,B,x_0$ (as well as the corresponding parameters for $\varphi_{BS}(x)$) for different values of $Z$ can be found in \cite{Jablonski1992ApproximationOT}. Finding the Fourier image $\tilde{U}_{TFDJ}$ according to (\ref{eq3}) is not difficult (although somewhat tedious). Therefore, we do not present the (cumbersome) result used for numerical calculations.

Firsov proposed a very interesting one-parameter approximation of the Thomas-Fermi-Dirac screening function \cite{Firsov1982TheST}, which can be considered as a generalization of the Tietz approximation (\ref{eq35}) to the case of the Thomas-Fermi-Dirac model:
\begin{equation}
\varphi_{TFDF}(x)=\frac{\sinh^2{(\beta c)}}{\sinh^2{[\beta (x+c)]}},
\label{eq40}
\end{equation}
where 
\begin{equation}
\beta=\frac{1}{2}\left(\frac{81}{32\pi^2Z^2}\right )^{1/6},\;\;
c=\frac{1}{\beta}\ln{\left (a_F\beta+\sqrt{1+a_F^2\beta^2}\right )},
\label{eq41}
\end{equation}
and $a_F\approx 1.82$. Since \cite{Firsov1982TheST} is not an easy reading, in appendix \ref{AppB} we provide a brief derivation of $\varphi_{TFDF}(x)$ and its Fourier image. Unlike other Thomas-Fermi-Dirac screening functions and similarly to the Thomas-Fermi ones, Firsov approximation posses an universal scaling in $Z$. 

The third group of tested potentials includes analytical approximations of the numerical solution of self-consistent field models \cite{Jackiw} of either the Hartree-Fock type or the Dirac-Hartree-Fock-Slater type.

Strand and Bonham approximate analytical expressions for the Hartree-Fock potential of neutral atoms up to $Z=36$ as follows (note that
$\gamma_1^{(a)}+\gamma_2^{(a)}=1$) \cite{Strand1964AnalyticalEF}:
\begin{equation}
\varphi_{HFSB}(r)=\sum\limits_{i=1}^2\gamma_i^{(a)}e^{-\lambda_i^{(a)}r}+r\sum\limits_{i=1}^m\gamma_i^{(b)}e^{-\lambda_i^{(b)}r},
\label{eq42}
\end{equation}
where $m=2$ for $Z=2$ to $Z=18$ and $m=3$ for $Z=19$ to $Z=36$. The Fourier image of the corresponding potential has the form
\begin{equation}
\tilde{U}_{HFSB}(\tilde{q})=4\pi Z\sqrt{\alpha}a_B^2\left [\sum\limits_{i=1}^2\frac{\gamma_i^{(a)}}{\tilde{q}^2+\alpha_i^2}+\frac{4m_e}{m_\mu}\sum\limits_{i=1}^m\frac{\gamma_i^{(b)}a_0\beta_i}{(\tilde{q}^2+\beta_i^2)^2}\right ],
\label{eq43}
\end{equation}
where
\begin{equation}
    \alpha_i=\frac{2m_e}{m_\mu}\lambda_i^{(a)}a_0,\;\;
    \beta_i=\frac{2m_e}{m_\mu}\lambda_i^{(b)}a_0.
    \label{eq44}
\end{equation}
Since a potential consisting only of Yukawa terms is easier to use, Cox and Bonham approximated the screening function by a sum of only exponential functions \cite{Cox_1967}, such as the first term in (\ref{eq42}). For atoms from $Z=1$ to $Z=21$, they use Hartree-Fock non-relativistic wave functions, while for the atoms from $Z=22$ to $Z=54$ Dirac-Hartree-Fock-Slater relativistic  wave functions were used, in which the one-electron orbitals are solutions of the Dirac equation.

Even for intermediate atomic numbers, distortions in atomic potential due to relativistic effects are noticeable. Salvat et al. \cite{Salvat1987AnalyticalDS} used Dirac-Hartree-Fock-Slater relativistic  wave functions to provide reliable screening functions for all atoms $Z=1\div 92$. They approximate their numerical results by Moli\'{e}re-type analytical function, which facilitates their implementation.

Further results of calculating cross sections and their dependence on the type of approximation of the self-consistent potential are presented in the table \ref{Tab2}.

\begin{table}[H]
\begin{center}
\scriptsize
\begin{tabular}{|c|c|c|c|c|c|c|c|}
\hline 
$Z$ & \thead{Cross\\ Sections $[\mathrm{cm}^{2}]$} & \thead{Cox \\ \& Bonham } & \thead{Salvat \\ et al.} & Jablonski & \thead{Bonham \\ \& Strand} & \thead{Strand \\ \& Bonham} & Firsov
\tabularnewline
\hline 
\hline 
\multirow{2}{*}{$4$} & $\sigma(100\rightarrow211)\cdot 10^{23}$ & $5.006$  & $5.390$ & $5.109$ & $5.109$ & $5.276$  & $5.293$ \\
\cline{2-8} 
& $\sigma(100\rightarrow X)\cdot 10^{22}$ & $2.078$   & $2.215$ & $2.113$ & $2.112$ & $2.174$ & $2.180$
\tabularnewline
\hline 
\hline
\multirow{2}{*}{$13$} & $\sigma(100\rightarrow211)\cdot 10^{22}$ & $4.783$  & $4.822$ & $4.885$ & $4.892$ & $4.894$ & $5.046$
\tabularnewline
\cline{2-8} 
& $\sigma(100\rightarrow X)\cdot 10^{21}$ & $2.009$  & $2.024$ & $2.046$ & $2.049$ & $2.050$  & $2.105$
\tabularnewline
\hline 
\hline
\multirow{2}{*}{$82$} & $\sigma(100\rightarrow211)\cdot 10^{20}$ & $-$ & $1.511$ & $1.586$ & $1.591$ & $-$ & $1.637$
\tabularnewline
\cline{2-8} 
& $\sigma(100\rightarrow X)\cdot 10^{20}$ & $-$ & $6.563$ & $6.837$ & $6.856$ & $-$  & $7.034$
\tabularnewline
\hline 
\end{tabular}
\end{center}
\caption{The transition cross section $\sigma(100\rightarrow211)$
and the total cross section $\sigma(100\rightarrow X)$ for different approximations to the Thomas-Fermi-Dirac and self-consistent potentials.}
\label{Tab2}
\end{table}
As for the cross sections, the relative difference in the results of various approximations of the atomic potential can reach 8\%. The difference is significantly reduced in the cross-sectional ratios. For example, the difference in $\sigma(100\rightarrow211)/\sigma(100\rightarrow X)$ does not exceed 1\%. This decrease in difference is further illustrated for the average values in table \ref{Tab3}. This reduction plays a major role in explaining the fact that the yields of dimuonium quantum states (the solutions of transport equations (\ref{eq15}) which are dependent not directly on cross sections, but on the cross-sectional ratios) will be very similar for different approximations of the atomic potential.

\begin{table}[H]
\begin{center}
\small
\begin{tabular}{|c|c|c|c|}
\hline 
$Z$ & $\sigma_{av}(100\rightarrow 211)$ & $\sigma_{av}(100\rightarrow X)$ & $10\cdot\left (\frac{\sigma(100\rightarrow 211)}{\sigma(100\rightarrow X)}\right )_{av}$ \\[1ex]
\hline 
\hline 
$4$ & $(5.28\pm 0.13)\cdot 10^{-23}~\mathrm{cm}^2$ & $(2.175\pm 0.046)\cdot 10^{-22}~\mathrm{cm}^2$  &  $2.428\pm 0.008$   \\
\hline
$13$ & $(4.956\pm 0.093)\cdot 10^{-22}~\mathrm{cm}^2$  & $(2.072\pm 0.034)\cdot 10^{-21}~\mathrm{cm}^2$  & $2.392\pm 0.006$   \\
\hline
$82$ & $(1.597\pm 0.032)\cdot 10^{-20}~\mathrm{cm}^2$ & $(6.88\pm 0.12)\cdot 10^{-20}~\mathrm{cm}^2$  & $2.321\pm 0.007$   \\
\hline 
\end{tabular}
\end{center}
\caption{Average values of the transition cross section $\sigma(100\rightarrow211)$, the total cross section $\sigma(100\rightarrow X)$, and the average value of their ratio, for
different approximations to the Thomas-Fermi, Thomas-Fermi-Dirac and self-consistent potentials.}
\label{Tab3}
\end{table}

\section{Results on the yields of low-lying quantum states of dimuonium}
The results presented in this section concern the yields of low-lying dimuonium quantum states ($2P$, $2S$, $3P$) because of their importance and because their yields form the main contributions after the $1S$ quantum state, however, since equation (\ref{eq15}) describes the evolution of any $(n,l,m)$ dimuonium quantum state, it allows one to find yields of higher dimuonium quantum states as well, in practice up to about $n_{max}=10$. The yield for a state $(n,l)$ is obtained by summing the yields of the corresponding states $(n,l,m)$ over all their magnetic quantum numbers $m$. The quantization axis is along the beam direction unless otherwise indicated.

\begin{figure}[H]
\begin{center}
\includegraphics[scale=0.95]{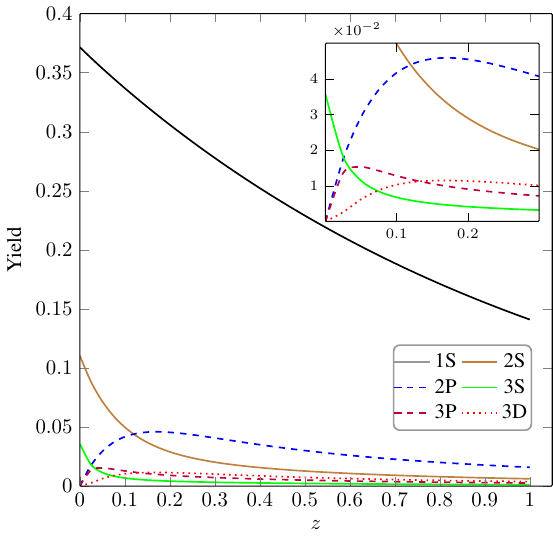}
\end{center}
\caption{Yields of low-lying dimuonium quantum states through the foil, expressed as fractional ratios of $N_{\alpha}$ to $N_{0}$, the number of initially produced dimuonium $1S$ states at the electron-positron interaction point. $1S$ -- black (upper) solid line, $2S$ -- brown (middle) solid line, $3S$ -- green (lower) solid line, $2P$ -- blue (upper) dashed line, $3P$ -- purple (lower) dashed line, $3D$ -- red dotted line. For $Z=13$, Thomas-Fermi-Moli\'{e}re atomic potential, distance from dimuonium production point to the foil $d=2~\mathrm{mm}$, and $n_{max}=5$.}
\label{fig_all_states}
\end{figure}

We have studied dependence of the yields of low-lying dimuonium quantum states on the following: truncation of the infinite system of transport equations to a finite number of quantum states, target material, inclusion of natural decay channel in the transport equation, and the choice of quantization axis (along beam direction or transferred momentum). Fig.\ref{fig_all_states} illustrates the relative magnitudes of the yields of low-lying dimuonium quantum states when it passes through aluminium foil, expressed as fractional ratios of $N_{\alpha}$ to $N_{0}$, the number of initially produced dimuonium $1S$ states at the electron-positron interaction point. Numerical results from which Fig.\ref{fig_all_states} was prepared show that the yields of the in-foil produced low-lying dimuonium quantum states, mainly $2P$ and $3P$, when the  distance from dimuonium production point to the foil is $d=2mm$, can reach $12\%$ and $3\%$ respectively, compared to the main input to the foil which is the $1S$ state. 

\begin{figure}[H]
\begin{center}
\includegraphics[scale=0.9]{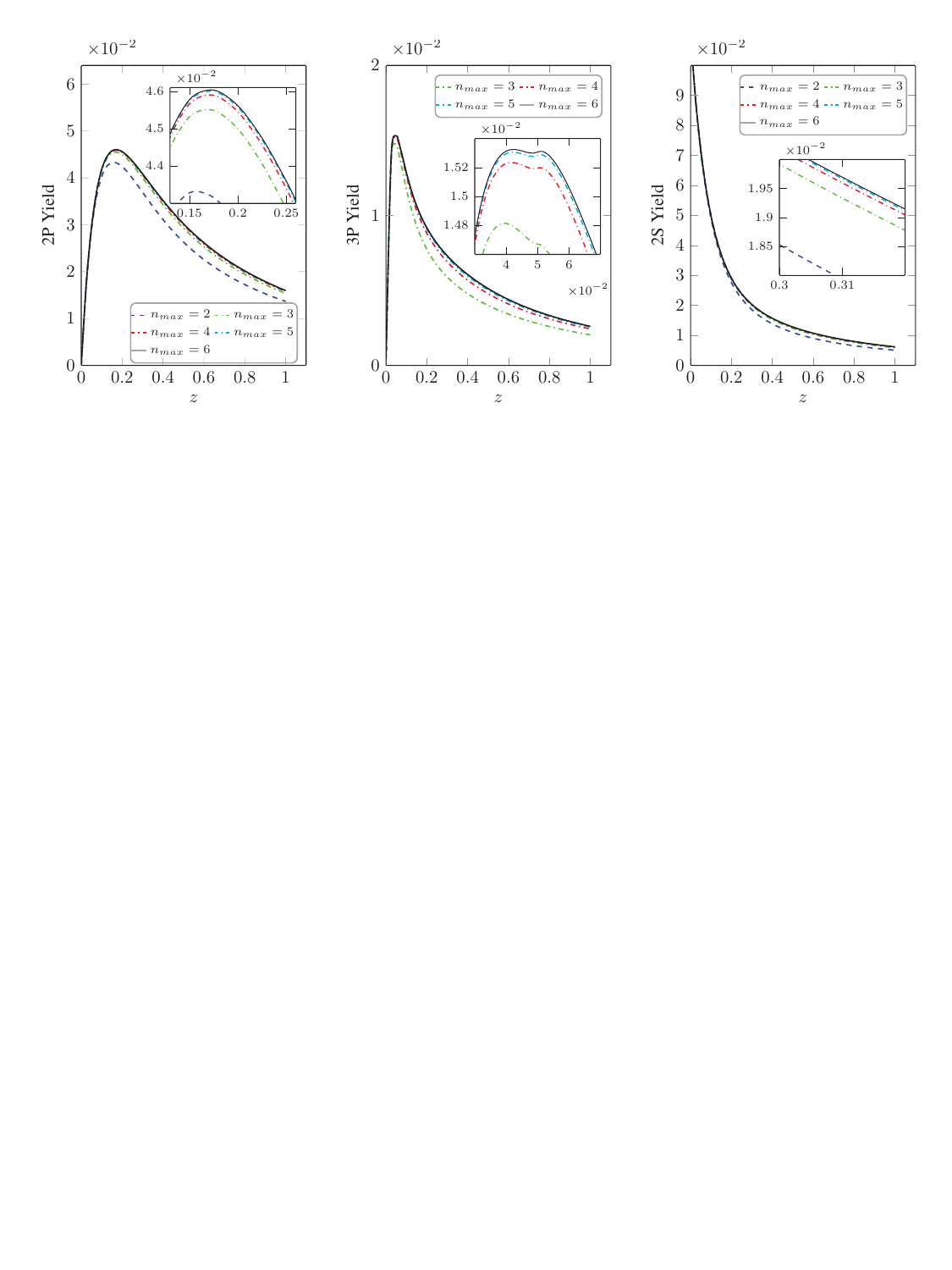}
\end{center}
\caption{Convergence with respect to the truncation parameter $n_{max}$ for the yields of low-lying dimuonium quantum states: $2P$ left figure, $3P$ middle figure, and $2S$ right figure. $n_{max}=2$ -- blue dashed line, $n_{max}=3$ -- green (lower) dash-dotted line, $n_{max}=4$ -- red (middle) dash-dotted line, $n_{max}=5$ -- cyan (upper) dash-dotted line, $n_{max}=6$ -- black solid line. For $Z=13$, Thomas-Fermi-Moli\'{e}re atomic potential, distance from dimuonium production point to the foil $d=2~\mathrm{mm}$. The insets enhance the visibility for a clear distinction between the convergent curves.}
\label{fig1}
\end{figure}

The first question we have investigated is how the yields of dimuonium low-lying states depend on the finite-reduction of the transport equations. We varied the truncation parameter $n_{max}$ from 2, which corresponds to a total of 5 active quantum states $(n,l,m)$ in the transport equations, to 10 (a total of 385 active states in the transport equations). Fig.\ref{fig1} illustrates the results obtained. As can be seen, the results indicate fairly good convergence in $n_{max}$: they do not change appreciably already after $n_{max}=5$. Convergence was checked for all atomic potential approximations used.

\begin{figure}[H]
\begin{center}
\includegraphics[scale=0.9]{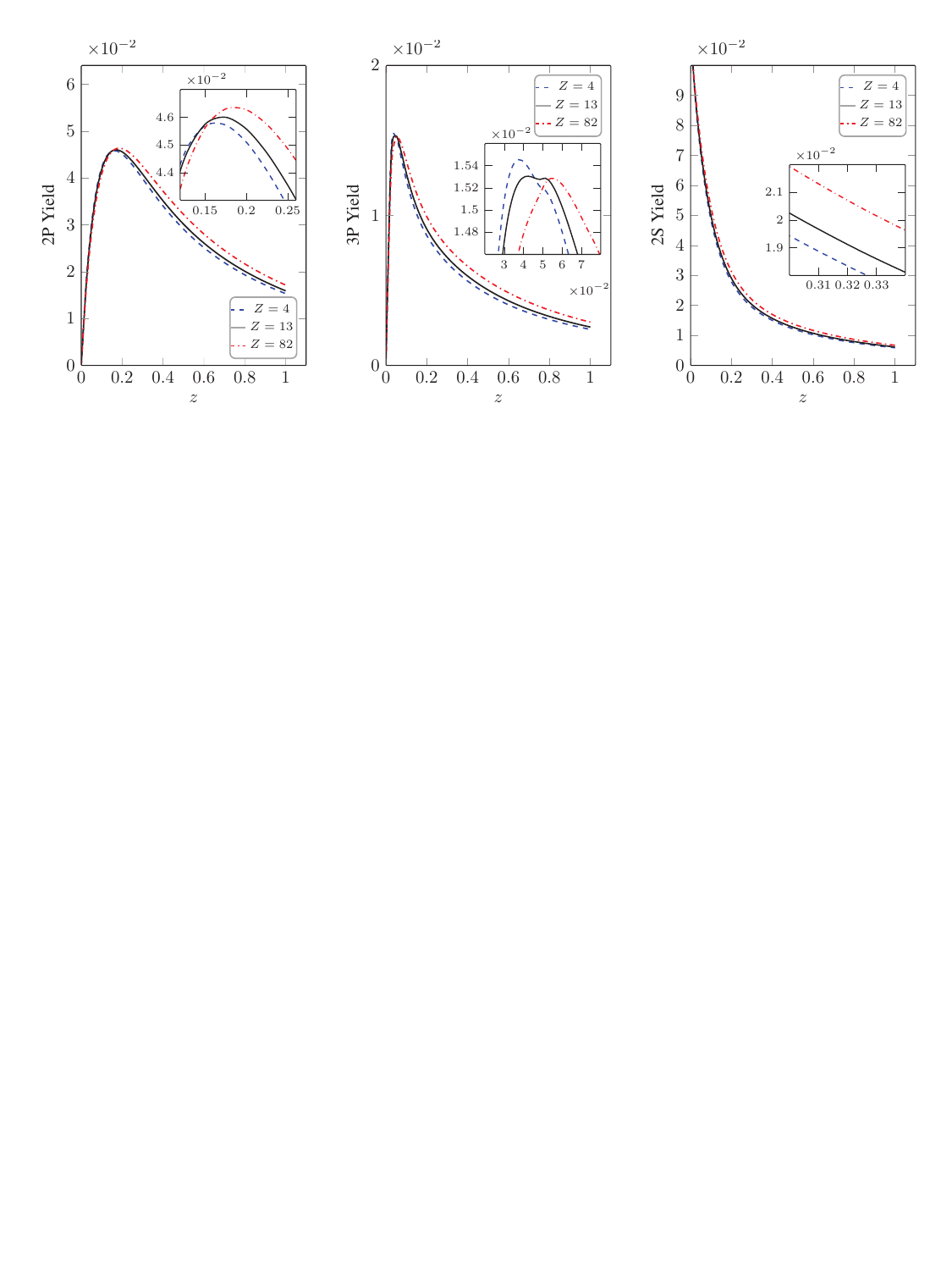}
\end{center}
\caption{Dependence of the yields of low-lying dimuonium quantum states on target material. From left to right: $2P$, $3P$, $2S$. Beryllium -- blue dashed (lower) line, Aluminium -- black solid line, Lead -- red dash-dotted (upper) line. For Thomas-Fermi-Moli\'{e}re atomic potential, distance to foil $d=2~\mathrm{mm}$, $n_{max}=5$.}
\label{fig2}
\end{figure}

As already mentioned, transport equations in dimensionless form (\ref{eq15}) give results that are little sensitive to the target material. This is illustrated in Fig.\ref{fig2}. 

\begin{figure}[H]
\begin{center}
\includegraphics[scale=0.9]{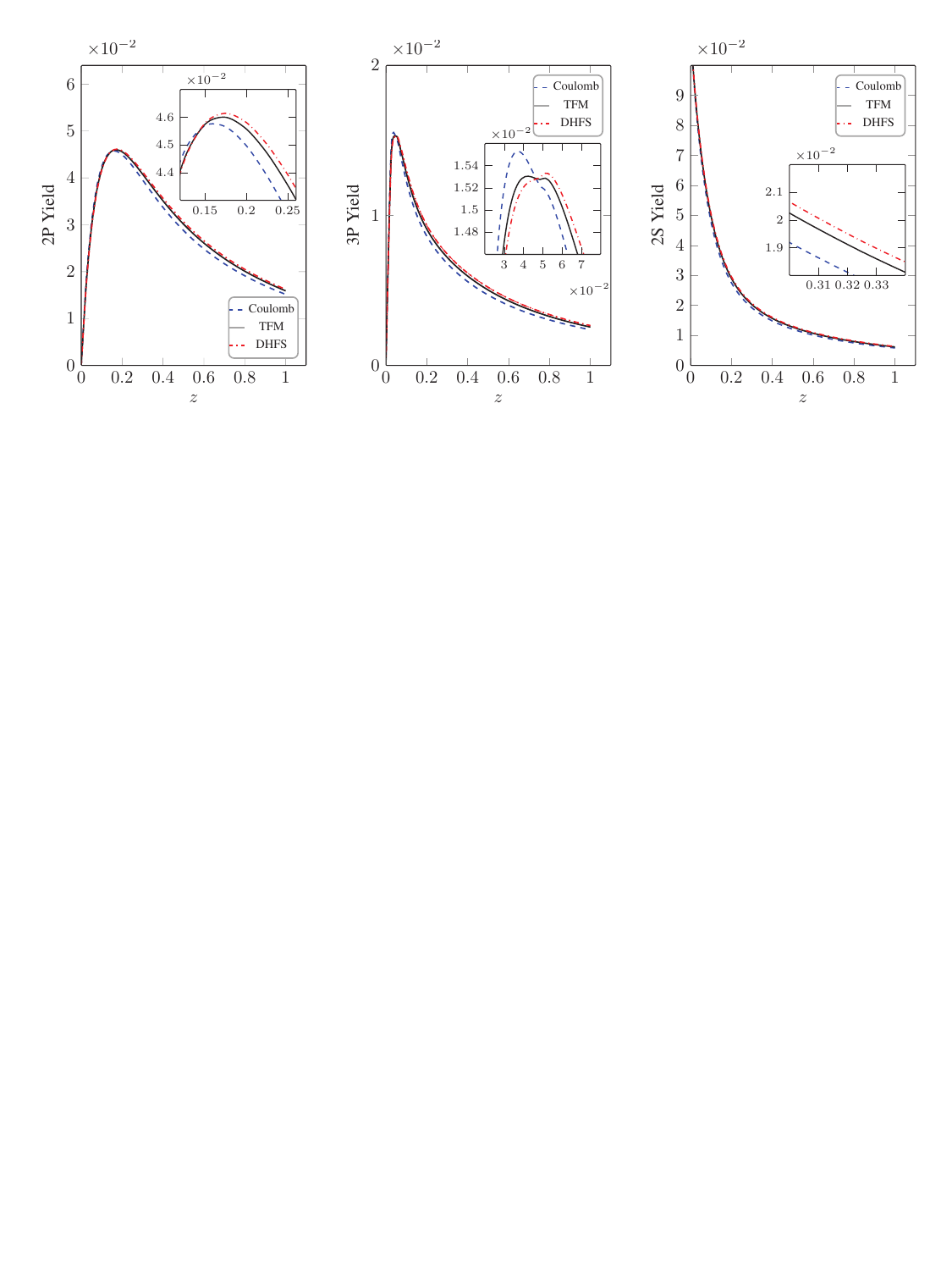}
\end{center}
\caption{Dependence of the yields of low-lying  dimuonium quantum states on the used approximation of the atomic potential. From left to right: $2P$, $3P$, $2S$. Truncated Coulomb -- blue dashed (lower) line, Thomas-Fermi-Moli\'{e}re -- black solid line, Dirac-Hartree-Fock-Slater -- red dash-dotted (upper) line. Other approximations' results lie between DHFS and Truncated Coulomb. $Z=13$, distance to foil $d=2~\mathrm{mm}$, $n_{max}=5$.}
\label{fig3}
\end{figure}

The dependence of the results obtained for the yields of low-lying quantum states of dimuonium on the atomic potential approximation used is illustrated in Fig.\ref{fig3}. We found that all studied approximations give very similar yield results. As mentioned before, this can be explained by the fact that the cross-sectional ratios vary very slightly for different approximations of the atomic potential, as shown in table \ref{Tab3}. Therefore, beside the results for the Thomas-Fermi-Moli\'{e}re atomic potential, the results for only two other potentials, Dirac-Hartree-Fock-Slater \cite{Salvat1987AnalyticalDS} and truncated Coulomb (described in next section), which produce the maximal differences from the Thomas-Fermi-Moli\'{e}re case, are illustrated in Fig.\ref{fig3}.

\begin{figure}[H]
\begin{center}
\includegraphics[scale=0.9]{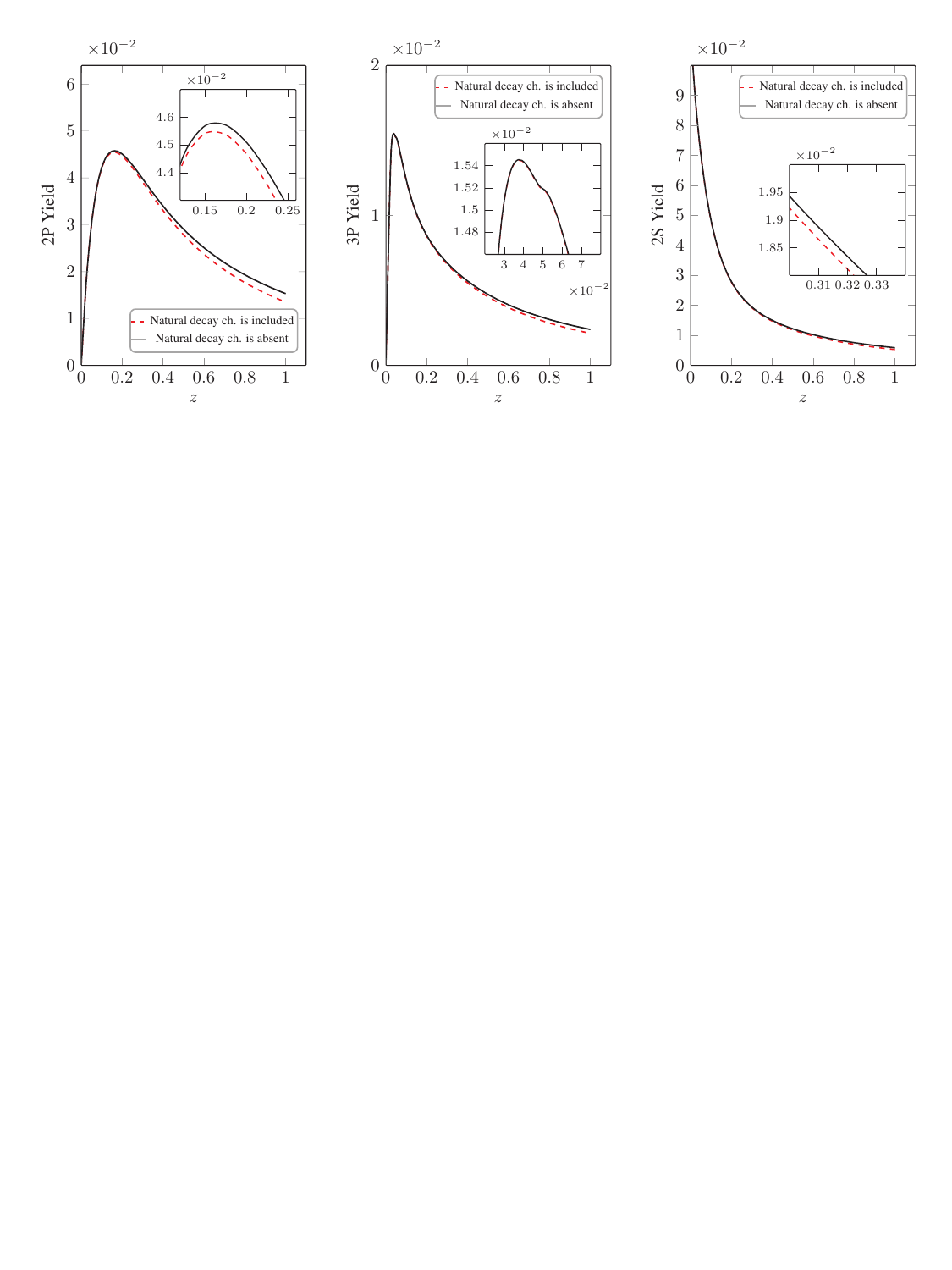}
\end{center}
\caption{Dependence of the yields of low-lying  dimuonium quantum states on the presence of the natural decay channel $(\mu^+\mu^-)\to e^+e^-$, for beryllium $(Z=4)$. From left to right: $2P$, $3P$, $2S$. Black solid line -- the natural decay channel $(\mu^+\mu^-)\to e^+e^-$ is absent. Red dashed line -- the natural decay channel was included in the transport equations for the $1S$ and $2S$ states. Thomas-Fermi-Moli\'{e}re atomic potential, distance to foil $d=2~\mathrm{mm}$, $n_{max}=5$.}
\label{fig4}
\end{figure}

As already mentioned, the $(\mu^+\mu^-)\to e^+e^-$ natural decay channel was not included in the transport equation. The contribution of this channel is relatively important only for low-$Z$ materials. Its magnitude is illustrated in Fig.\ref{fig4} for beryllium ($Z=4$) and for the Thomas-Fermi-Moli\'{e}re potential.

\begin{figure}[H]
\begin{center}
\includegraphics[scale=0.9]{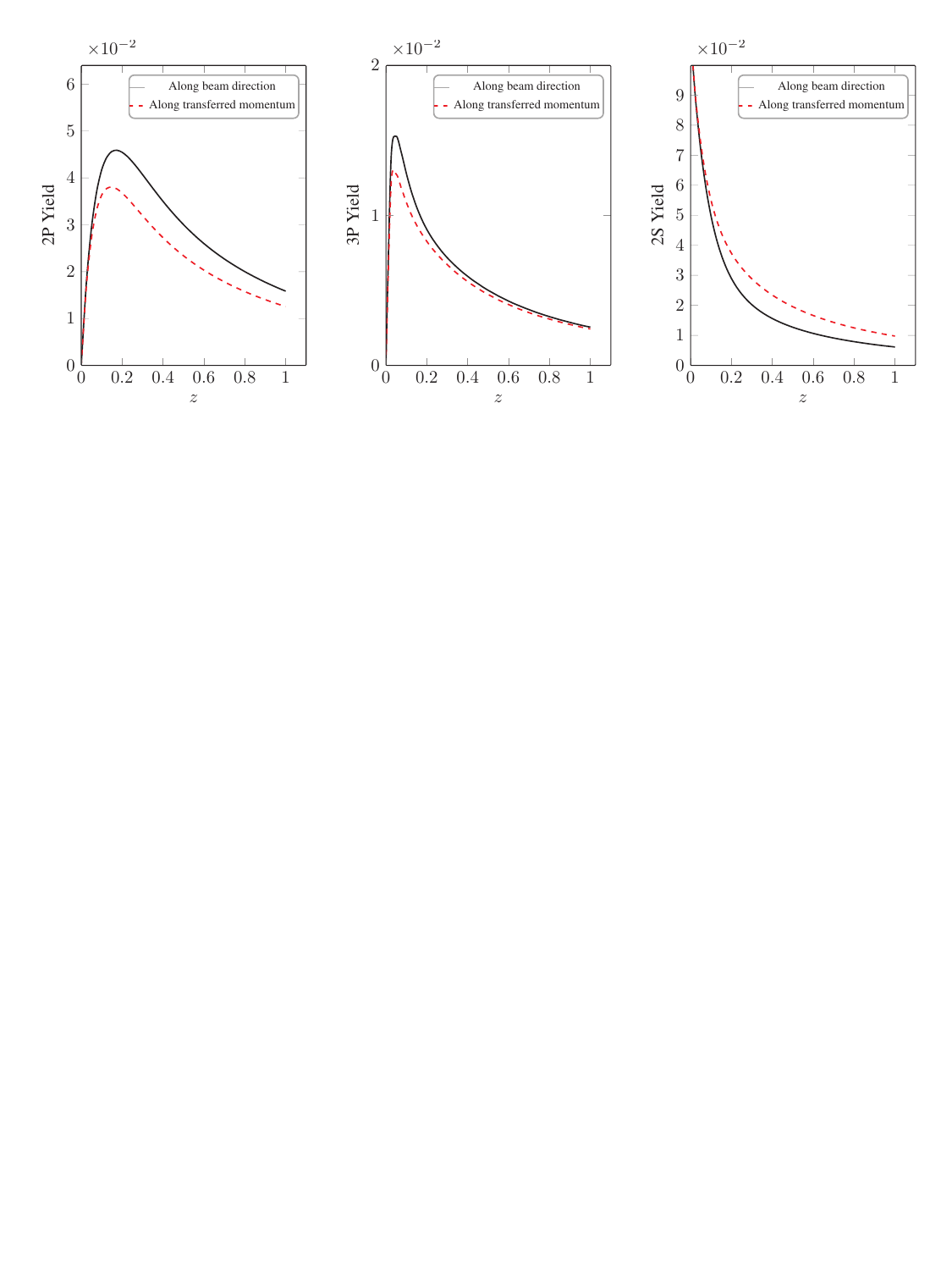}
\end{center}
\caption{Comparison of the yields of low-lying dimuonium quantum states based on the choice of the quantization axis. From left to right: $2P$, $3P$, $2S$. Black solid line -- the quantization axis is along beam direction. Red dashed line -- the quantization axis is along transferred momentum. $Z=13$, Thomas-Fermi-Moli\'{e}re atomic potential, distance to foil $d=2~\mathrm{mm}$, $n_{max}=5$.}
\label{fig_quantization}
\end{figure}

The last important result, which will be shown in this section, is based on a comparison of the yields of low-lying dimuonium quantum  states for various options for choosing the quantization axis when calculating cross sections. The cross-section results summed over magnetic quantum numbers should not depend on the choice of the quantization axis. More precisely, we must have
\begin{equation}
\sum\limits_{m_1,m_2}\sigma(n_1,l_1,m_1\to n_2,l_2,m_2)=
\sum\limits_{m_1^\prime,m_2^\prime}\sigma^\prime(n_1,l_1,m_1^\prime\to n_2,l_2,m_2^\prime).
\label{eq_axis} 
\end{equation}
However, it is clear that the choice of the quantization axis along the transferred momentum is incompatible with the transport equations used in this work, since the direction of the transferred momentum varies from scattering to scattering. Therefore, we will get incorrect results for the yields for this choice, even though the yields are summed over the magnetic quantum numbers. Nevertheless, it is interesting to know how much we err when we make the wrong choice of the quantization axis along the transferred momentum to calculate the cross sections used in the transport equations (\ref{eq15}), since such a choice was made, for example, in \cite{Banburski:2012tk} (for $n_{max}=2$). Our results indicate that the errors are not dramatic but somewhat significant, as illustrated in Fig.\ref{fig_quantization}.

\section{The importance of screening in atomic potential}
A somewhat unexpected result of our study was the very low sensitivity of the results obtained to the approximation of the atomic potential used. The following reasoning partially explains this low sensitivity\footnote{We owe this insight to a comment made to us by Alexander Milstein.}. The energy scale in the dimuonim spectra is $\frac{1}{2}m_{\mu}\alpha^{2}$. Therefore, for dimuonium to be excited, the required transferred momentum must satisfy $\frac{q^{2}}{2m_{\mu}}\sim\frac{1}{2}m_{\mu}\alpha^{2}$, and the characteristic (dimensionless) momentum for excitation is $\tilde{q}=qa_B\sim \alpha m_\mu a_B=2$. On the other hand,
the transferred momentum corresponding to the Thomas-Fermi screening scale $b$ is $\tilde{q}_c=\frac{1}{b}a_B\approx 1.09\cdot 10^{-2}Z^{1/3}$. Even for $Z=82$, $\tilde{q}_c\approx5\cdot 10^{-2}\ll 2$. 
In other words, dimuonium must pass very close to the nucleus to be excited, and this is the region where most of the scattering takes place. The screening effect in this region very close to the nucleus is small, so it appears that the dimuonium scattering is determined mainly by the bare Coulomb potential of the nucleus.  

But it is impossible to completely remove the screening effect, since the corresponding integrals in the case of a pure Coulomb potential will diverge logarithmically. For example, for $nS$ states we have \cite{Afanasyev_1993,Alizzi2021AlternativeIO}
\begin{equation}
F_{n00}^{n00}=\frac{2}{n(4+n^2\tilde{q}^2)}\,U_{n-1}\left(\frac{4-n^2\tilde{q}^2}{4+n^2\tilde{q}^2}\right) \left [ P_{n-1}\left(\frac{4-n^2\tilde{q}^2}{4+n^2\tilde{q}^2}\right)+ P_n\left(\frac{4-n^2\tilde{q}^2}{4+n^2\tilde{q}^2}\right)\right ],
\label{eq45}
\end{equation}
where $P_n(x)$ are Legendre polynomials, and $U_n(x)$ are Chebyshev polynomials of the second kind. Using (the prime denotes derivative with respect to $x$)
\begin{equation}
P_n(1)=1,\;\;P_n^\prime(1)=\frac{1}{2}n(n+1),\;\;U_n(1)=n+1,\;\;U_n^\prime(1)=\frac{1}{3}n(n+1)(n+2),
\label{eq46}
\end{equation}
we get for small $\tilde{q}$
\begin{equation}
1-F_{n00}^{n00}\approx \frac{1}{12}n^2(5n^2+1)\,\tilde{q}^2.
\label{eq47}
\end{equation}
Then, for the pure Coulomb potential $\tilde{U}(q)=4\pi Ze/q^2$, equation (\ref{eq2}) gives
\begin{equation}
\sigma_{n00}^{tot}\approx \frac{4\pi\alpha^2Z^2}{3V^2}\,n^2(5n^2+1)\,a_B^2\int\frac{dq}{q}.
\label{eq48}
\end{equation}
Integration limits in this logarithmically divergent integral can be taken from the inverse screening length $1/b$ to the inverse size of the $(n,0,0)$ quantum state of dimuonium $1/(n^2a_B)$, and we get in light of (\ref{eq21})
\begin{eqnarray} &&
\sigma_{n00}^{tot}\approx \frac{4\pi\alpha^2Z^2}{3V^2}\,n^2(5n^2+1)\,a_B^2\ln{\left[\frac{m_\mu}{2n^2m_e}\left(\frac{9\pi^2}{128 Z}\right)^{1/3}\right ]}\approx \nonumber \\ &&
\frac{4\pi\alpha^2Z^2}{3V^2}\,n^2(5n^2+1)\,a_B^2\left [4.5167-\frac{1}{3}Z-2\ln{n}\right ].
\label{eq49}
\end{eqnarray}
We expect (\ref{eq49}) to be a good approximation if $b\gg n^2a_B$, which translates to
\begin{equation}
n\ll\sqrt{\frac{b}{a_B}}\approx \frac{9.57}{Z^{1/6}}.
\label{eq50}
\end{equation}
Compared to the results for the Thomas-Fermi-Moli\'{e}re potential, (\ref{eq49}) provides accuracy better than 5\% for $n =1$ for all elements $Z=1\div 98$ and for $n=2$ up to $Z=28$. For $Z=28 \div 98$, the accuracy is better than 10\% for $n=2$, and it is better than 30\% for $n=3$ for all elements $Z=1 \div 98$.

Perhaps the last equality in (\ref{eq46}) needs some comment, and this is done in appendix \ref{AppC}.

\begin{figure}[H]
\begin{center}
\includegraphics[scale=0.9]{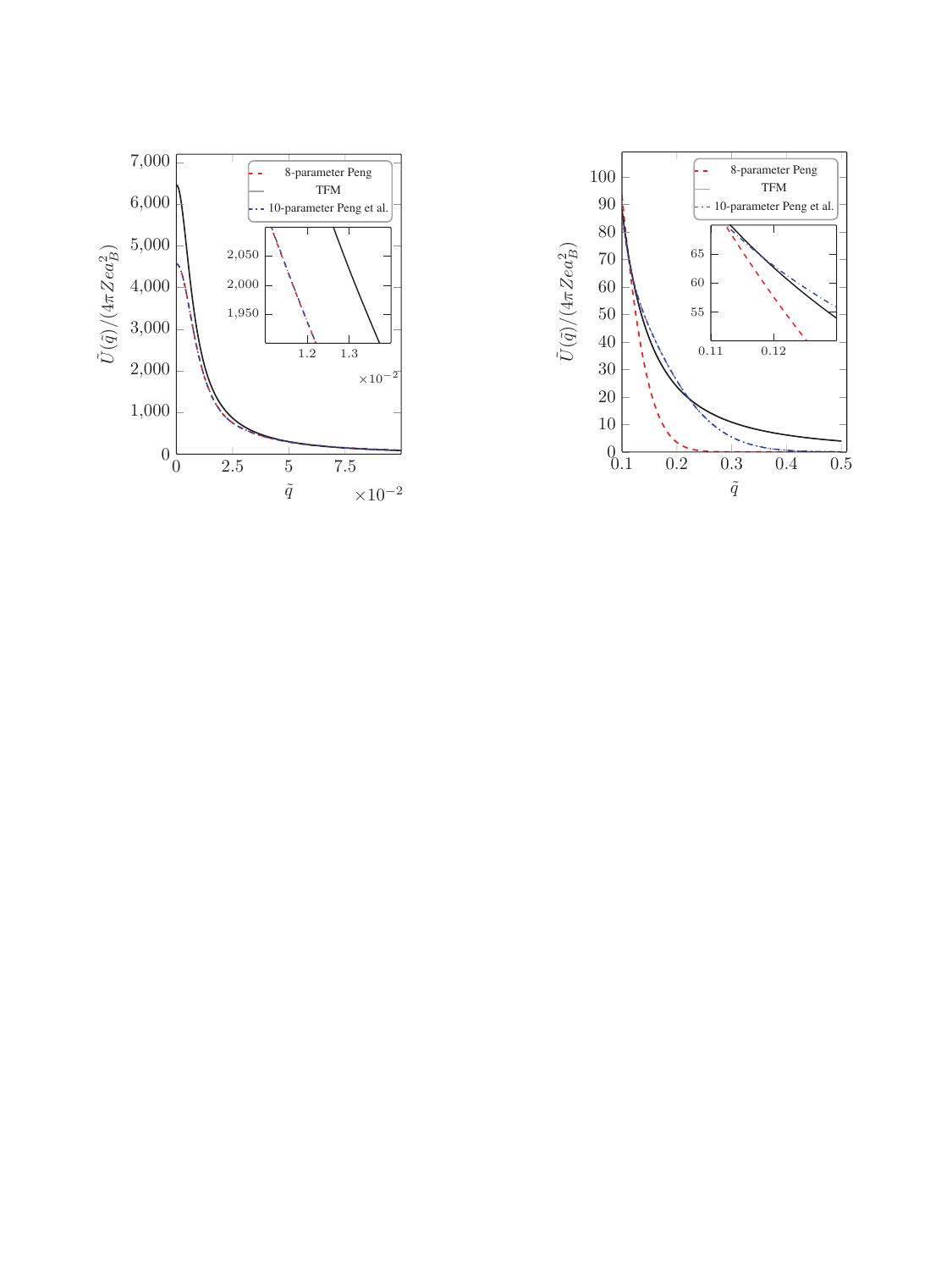}
\end{center}
\caption{Comparison of dimensionless Fourier images of potentials reconstructed from \cite{Peng1996RobustPO} and \cite{Peng1999ELECTRONAS} data with the Fourier image of the Thomas-Fermi-Moli\'{e}re potential. Black solid line -- Thomas-Fermi-Moli\'{e}re. Red dashed line -- Eight-parameter Peng  model \cite{Peng1999ELECTRONAS}. Blue dash-dotted line -- Ten-parameter Peng et al. model \cite{Peng1996RobustPO}. For $Z=13$.}
\label{fig5}
\end{figure}

In order to additionally check the idea that the main part of dimuonium transitions is determined by the pure Coulomb field of the nucleus with a screening that has a small effect, we examined three more models for the atomic potential:
\begin{itemize}
\item Truncated Coulomb: if $q<\frac{1}{b}$, $\tilde{U}(q)=0$, and
if $q>\frac{1}{b}$, $\tilde{U}(q)=4\pi Ze/q^2$. 
\item Eight-parameter Peng  model \cite{Peng1999ELECTRONAS} and Coulomb: if $s<2\text{\AA}^{-1}$,
Peng model potential, and if $s>2\text{\AA}^{-1}$, Coulomb potential ($q=4\pi s$).
\item Ten-parameter Peng et al. model \cite{Peng1996RobustPO} and Coulomb: if $s<6\text{\AA}^{-1}$,
Peng et al. model potential, and if $s>6\text{\AA}^{-1}$, Coulomb potential.
\end{itemize}
In \cite{Peng1996RobustPO,Peng1999ELECTRONAS} they provide not the screening function $\varphi(r)$, but the electron atomic scattering factor
\begin{equation}
f^{(e)}(s)=\frac{2Ze}{a_0 \,q}\int\limits_0^\infty \varphi(r)\sin{(qr)}\,dr,\;\;q=4\pi s.
\label{eq51}
\end{equation}
Comparing with (\ref{eq3}), we see that the Fourier image of the corresponding potential is
\begin{equation}
\tilde{U}(q)=2\pi a_0f^{(e)}(s).
\label{eq52}
\end{equation}
To use (\ref{eq52}), it should be remembered that \cite{Peng1996RobustPO,Peng1999ELECTRONAS} provide $f^{(e)}(s)$ in $\text{\AA}$, $s$ in $\text{\AA}^{-1}$, and in our programs we use the dimensionless quantity $\tilde{U}(q)/a_B^2$. Fig.\ref{fig5} shows a comparison of the dimensionless potentials $\tilde{U}(q)/a_B^2$ as functions of $\tilde{q}=qa_B$, reconstructed from the data of \cite{Peng1996RobustPO} and \cite{Peng1999ELECTRONAS} with the Fourier image of the Thomas-Fermi-Moli\'{e}re potential. It is important to remember that the numerical values of the electron atomic scattering factor obtained using relativistic Hartree–Fock atomic wave functions were approximated by the sum of four Gaussians in \cite{Peng1999ELECTRONAS} and the sum of five Gaussians in \cite{Peng1996RobustPO}. Accordingly, these results apply only to the ranges $s=0\div 2\text{\AA}^{-1}$ ($\tilde{q}=0\div 0.129$) and $s=0\div 6 \text {\AA}^{-1}$ ($\tilde{q}=0\div 0.386$) respectively, where the fitting was performed. Any attempts to use them outside of these ranges may result in large errors.

The results summarized in Tab.\ref{Tab4} show that the nucleus Coulomb potential indeed makes the main contribution to the scattering of dimuonium on a target nucleus. At that, the truncated Coulomb potential
(with the region around of the zero-transferred-momentum singularity removed) gives results that are most different from the case of the Thomas-Fermi-Moli\'{e}re potential. Even in this case, the difference in the yields of low-lying states of dimuonium, according to Fig.\ref{fig3}, is insignificant. If a milder regularization of the Coulomb singularity is used with \cite{Peng1996RobustPO} or \cite{Peng1999ELECTRONAS} data, the results become closer to the Thomas-Fermi-Moli\'{e}re case.

\begin{table}[H]
\begin{center}
\scriptsize
\begin{tabular}{|c|c|c|c|c|c|}
\hline 
$Z$ & \thead{Cross\\ Sections $[\mathrm{cm}^{2}]$}  & \thead{Truncated \\Coulomb} & \thead{Coulomb \\ \& Peng8} & Moli\'{e}re & \thead{Coulomb \\ \& Peng10} \tabularnewline \hline 
\hline 
\multirow{2}{*}{$4$} & $\sigma(100\rightarrow211)\cdot 10^{23}$  & $6.248$ & $5.294$ & $5.351$ & $4.868 $   \\
\cline{2-6} 
& $\sigma(100\rightarrow X)\cdot 10^{22}$   & $2.525$ & $2.181$ & $2.200$ & $2.021$  \\
\cline{2-6}
& $\sigma(100\rightarrow211)/\sigma(100\rightarrow X)$ & $0.2474$ & $0.2427$ & $0.2432$ & $0.2409$ \\
\hline 
\hline
\multirow{2}{*}{$13$} & $\sigma(100\rightarrow211)\cdot 10^{22}$   & $5.965$ & $5.073$ & $5.013$ & $4.564$  \tabularnewline
\cline{2-6} 
& $\sigma(100\rightarrow X)\cdot 10^{21}$  & $2.438$ & $2.116$ & $2.093$ & $1.927$   \tabularnewline
\cline{2-6}
& $\sigma(100\rightarrow211)/\sigma(100\rightarrow X)$ & $0.2447$ & $0.2397$ & $0.2395$ & $0.2368$ \\
\hline 
\hline
\multirow{2}{*}{$82$} & $\sigma(100\rightarrow211)\cdot 10^{20}$ & $1.968$ & $1.736$ & $1.603$ & $1.578$   \tabularnewline
\cline{2-6} 
& $\sigma(100\rightarrow X)\cdot 10^{20}$ & $8.239$ & $7.400$ & $6.901$ & $6.821$  \tabularnewline
\cline{2-6}
& $\sigma(100\rightarrow211)/\sigma(100\rightarrow X)$ & $0.2389$ & $0.2346$ & $0.2323$ & $0.2313$ \\
\hline 
\end{tabular}
\end{center}
\caption{The transition cross section $\sigma(100\rightarrow211)$
and the total cross section $\sigma(100\rightarrow X)$ for the truncated Coulomb potentials of various types in  comparison with the Thomas-Fermi-Moli\'{e}re potential. The cross sections ratio is also shown.}
\label{Tab4}
\end{table}
Not surprisingly, the truncated Coulomb potential, which introduces a somewhat arbitrary cutoff parameter, gives the largest discrepancy with all previous results that take screening into account more accurately. We emphasize that the truncated Coulomb potentials were introduced only to understand the importance of screening effects, and the corresponding results should be discarded in any future reliable calculation of cross sections and yields.

\section{Concluding Remarks}
We hope that the results presented in this work are reliable, since as is described in previous sections, the calculation of cross sections are based on the MuMuPy calculator whose reliability was tested by three independent methods of implementation of atomic form factors, and the calculation of the yields are based on our computer code which was tested by two independent implementation methods for numerically solving truncated transport equations.

The results can be summarized as follows:
\begin{itemize}
    \item Based on three models of atomic potential (Thomas-Fermi, Thomas-Fermi-Dirac and self-consistent field), 15 different approximations were studied, including those taking into account relativistic effects in atomic orbitals. The relative difference in cross sections due to different approximations of the target atom potential (after excluding the truncated Coulomb potentials, which give the greatest discrepancy with all other results) can reach $8\%$. This fact reflects the relative importance of screening effects in the atomic potential, while the main contribution to the scattering of dimuonium on the target nucleus is determined by the bare Coulomb potential of the nucleus.
    \item The yields of the low-lying dimuonium quantum states obtained in the foil, mainly $2P$ and $3P$, when the distance from the dimuonium formation point to the foil is $d=2~mm$, can reach $12\%$ and $3\%$, respectively, compared to the main input to the foil, which is the $1S$ state. Since the yields will be different for different distances $d$, determining the optimal distance in the experiment will be a technical issue for experimenters.
    \item The results show good convergence of the yields for low-lying dimuonium states with respect to truncation of the infinite system of transport equations to a finite number of quantum states, and after $n_{max}=5$ the yields change only slightly.
    \item The transport equations in dimensionless form give results for yields that are insensitive to the target material (three different foils were tested: beryllium, aluminum, and lead). However, it is important to remember that after restoring the dimensional form of the transport equations, the real sensitivity to the foil material for a fixed foil thickness will be restored and will be significant for a large difference in $Z$, since the $Z^2$ coefficient in the cross-sections will be restored.
    \item The corresponding target atomic-potential-model-dependent error in the yields is quite small within the framework of the applied Born approximation, and for low-lying dimuonium quantum states it does not exceed $1\%$. 
    \item The contribution of the natural dimuonium decay channel to the transport equation for thin targets is relatively important only for low-$Z$ materials.
    \item The results show that significant errors can arise due to the choice of the quantization axis along the transferred momentum, which is incompatible with the transport equations used in this work.
\end{itemize}

\appendix
\section{Fourier image of the Robert's potential}
\label{AppA}
We need to calculate
\begin{equation}
\tilde U(\tilde{q})=\frac{4\pi Z e}{\tilde{q}}\,r_B^2\,\int\limits_0^\infty \sin{(\tilde{q}x)}\left (1+\bar{\eta}\sqrt{x}\right )e^{-\bar{\eta}\sqrt{x}}dx.
\label{A1}
\end{equation}
By changing the variables $x\to x^2/\tilde{q}$, the integral takes the form
\begin{equation}
\tilde U(\tilde{q})=\frac{8\pi Z e}{\tilde{q}^2}\,r_B^2\,\int\limits_0^\infty \sin{(x^2)}x\left (1+\eta x\right )e^{-\eta x}dx,
\label{A2}
\end{equation}
which can be rewritten as
\begin{equation}
\tilde U(\tilde{q})=\frac{4\pi Z e}{i\tilde{q}^2}\,r_B^2\,\left(\eta\frac{d^2}{d\eta^2}-\frac{d}{d\eta}\right)\left [I_+(\eta)-I_-(\eta)\right ],
\label{A3}
\end{equation}
with
\begin{equation}
I_\pm(\eta)=\int\limits_0^\infty e^{\pm ix^2-\eta x} dx.
\label{A4}
\end{equation}
Using integration by parts, we have 
\begin{equation}
I_+=-\frac{1}{\eta}\int\limits_0^\infty e^{ix^2}de^{-\eta x}=\frac{1}{\eta}+\frac{2i}{\eta}\int\limits_0^\infty xe^{ix^2-\eta x}dx=\frac{1}{\eta}-\frac{2i}{\eta}\frac{dI_+}{d\eta}.
\label{A5}
\end{equation}
This is a first order inhomogeneous differential equation for $I_+(\eta)$, and it can be solved by the method of variation of constants: substituting $I_+(\eta)=A(\eta)e^{i\eta^2/4}$ in (\ref{A5}), we obtain a simple homogeneous differential equation for $A(\eta)$, with a solution of the form
\begin{equation}
A(\eta)-A(0)=\frac{1}{2i}\int\limits_0^\eta e^{-ix^2/4}dx=\frac{1}{i}\int\limits_0^{\eta/2} e^{-ix^2}dx=
-i\left [\CF\left(\frac{\eta}{2}\right )-i\SF\left(\frac{\eta}{2}\right )\right ].
\label{A6}
\end{equation}
On the other hand
\begin{equation}
A(0)=I_+(0)=\int\limits_0^\infty e^{ix^2}dx=\sqrt{\frac{\pi}{8}}(1+i).
\label{A7}
\end{equation}
Therefore,
\begin{equation}
I_+=e^{i\eta^2/4}\left [(1+i)\sqrt{\frac{\pi}{8}}-i\CF\left(\frac{\eta}{2}\right )- \SF\left(\frac{\eta}{2}\right )\right ].
\label{A8}
\end{equation}
Similarly we get
\begin{equation}
I_-=e^{-i\eta^2/4}\left [(1-i)\sqrt{\frac{\pi}{8}}+i\CF\left(\frac{\eta}{2}\right )- \SF\left(\frac{\eta}{2}\right )\right ].
\label{A9}
\end{equation}
To calculate (\ref{A3}), it is useful to use equations
\begin{equation}
\frac{d}{d\eta}\left( I_+-I_-\right)=\frac{1}{2i}\left [2-\eta\left( I_++I_-\right)\right ],
\label{A10}
\end{equation}
and
\begin{equation}
\frac{d^2}{d\eta^2}\left( I_+-I_-\right)=-\frac{1}{2i}\left( I_++I_-\right)-\frac{\eta^2}{4}\left( I_+-I_-\right),
\label{A11}
\end{equation}
which follow from the differential equations for $I_+$ and $I_-$.

Now we have all the ingredients at hand to check the correctness of equation (\ref{eq31}).
\section{Firsov's potential}
\label{AppB}
The Thomas-Fermi-Dirac screening function obeys the differential equation \cite{Jackiw}
\begin{equation}
\varphi^{\prime\prime}=x\left [ \sqrt{\frac{\varphi}{x}}+\beta_0 \right ]^3, \;\;\;\beta_0=\left(\frac{3}{32\pi^2 Z^2}\right)^{1/3}.
\label{B1}
\end{equation}
Since $\beta_0$ is small, we can expand (\ref{B1}) with respect to it and replace (\ref{B1}) with the approximate equation
\begin{equation}
\varphi^{\prime\prime}=\frac{\varphi^{3/2}}{\sqrt{x}}+3\beta_0\varphi.
\label{B2}
\end{equation}
Firsov's main argument is as follows. The smallness of $\beta_0$ guarantees that the solution to the Thomas-Fermi-Dirac equation (\ref{B1}) will be close to the Thomas-Fermi screening function.
On the other hand, Tietz approximation (\ref{eq35}) of the Thomas-Fermi screening function, in addition to the Thomas-Fermi equation $\varphi_{Tietz}^{\prime\prime}\approx \frac{\varphi_{Tietz}^{3 /2}}{\sqrt{x}}$,
satisfies the equation
\begin{equation}
\varphi_{Tietz}^{\prime\prime}=\frac{6a_T^2}{(a_T+x)^4}=\frac{6\varphi_{Tietz}^2}{a_T^2}.
\label{B3}
\end{equation}
Therefore,
\begin{equation}
\frac{\varphi^{3/2}}{\sqrt{x}}\approx  \frac{\varphi_{Tietz}^{3/2}}{\sqrt{x}}\approx  \varphi_{Tietz}^{\prime\prime}=\frac{6\varphi_{Tietz}^2}{a_T^2}\approx  \frac{6\varphi^2}{a_F^2},
\label{B4}
\end{equation}
which allows us to replace equation (\ref{B2}) with
\begin{equation}
\varphi^{\prime\prime}=\frac{6\varphi^2}{a_F^2}+3\beta_0\varphi.
\label{B5}
\end{equation}
It is convenient to introduce a new parameter $\beta$ through
\begin{equation}
4\beta^2=3\beta_0
\label{B6}
\end{equation}
and rewrite (\ref{B5}) in the form
\begin{equation}
\varphi^{\prime\prime}=\frac{6\varphi^2}{a_F^2}+4\beta^2\varphi.
\label{B7}
\end{equation}
Multiplying both sides of (\ref{B7}) by $\varphi^\prime$ and integrating, assuming that $\varphi(x)$ and its derivatives vanish at infinity, we obtain the first integral
\begin{equation}
\varphi^{\prime\,2}=4\varphi^2\left (\frac{\varphi}{a_F^2}+\beta^2\right).
\label{B8}
\end{equation}
Therefore ($\varphi^\prime<0$ since the screening increases as $x$ increases, and correspondingly $\varphi$ decreases from one to zero), 
\begin{equation}
\varphi^\prime=-2\varphi\sqrt{\frac{\varphi}{a_F^2}+\beta^2}.
\label{B9}
\end{equation}
Separating the variables and integrating with the boundary condition $\varphi(0)=1$, we get
\begin{equation}
-2\frac{x}{a_F}=\int\limits_1^\varphi\frac{dt}{t\sqrt{t+a_F^2\beta^2}}=\frac{1}{a_F\beta}\ln{\left[\frac{\left(a_F\beta-\sqrt{\varphi+a_F^2\beta^2}\right)\left( a_F\beta+\sqrt{1+a_F^2\beta^2} \right)}
{\left(a_F\beta+\sqrt{\varphi+a_F^2\beta^2}\right)\left(a_F\beta-\sqrt{1+a_F^2\beta^2}\right)}\right ]}.
\label{B10}
\end{equation}
If a new parameter $c$ is introduced via
\begin{equation}
\frac{\sqrt{1+a_F^2\beta^2}-a_F\beta}{\sqrt{1+a_F^2\beta^2}+a_F\beta}=e^{-2\beta c},\;\;\mathrm{which\;is\;equivalent\;to}\;\;\sinh{\beta c}=a_F\beta,
\label{B11}
\end{equation}
the Firsov screening function is obtained from (\ref{B10}) in the compact form
\begin{equation}
\varphi(x)=\frac{a_F^2\beta^2}{\sinh^2{\beta(x+c)}}=\frac{\sinh^2{\beta c}}{\sinh^2{\beta(x+c)}}.
\label{B12}
\end{equation}
When $\beta c\ll 1$, $c\to a_F$ and $\sinh^2{\beta c}/\sinh^2{\beta(x+c)}\to a_F^2/(x+a_F)^2$. Therefore, the Firsov screening function (\ref{B12}) can be considered as a highly nontrivial generalization of the Tietz approximation from the Thomas-Fermi case to the Thomas-Fermi-Dirac case.

To find the Fourier image of the Firsov potential $Ze\varphi(x)/r,\;x=r/b$, we expand
\begin{equation}
\frac{1}{\sinh^2{x}}=\frac{4}{e^{2x}(1-e^{-2x})^2}=4\sum_{n=0}^\infty (n+1)e^{-2(n+1)x}=4\sum_{k=1}^\infty k e^{-2kx}.
\label{B13}
\end{equation}
Then
\begin{equation}
\int_0^\infty \frac{\sin{\bar{q}x}}{\sinh^2{\beta(x+c)}}\,dx=\bar{q}\sum_{k=1}^\infty\frac{ke^{-2k\beta c}}{k^2\beta^2+\bar{q}^2/4}=\bar{q}A(\bar{q}), \;\;\bar{q}=bq=\frac{b}{a_B}\tilde{q}.
\label{B14}
\end{equation}
Note that $A(\bar{q})$ can be rewritten in the following way 
\begin{eqnarray}  &&
A(\bar{q})=\frac{1}{\beta^2}\sum_{k=1}^\infty \frac{e^{-2k\beta c}}{k}-\frac{\bar{q}^2}{4\beta^2}\sum_{k=1}^\infty \frac{e^{-2k\beta c}}{k(k^2\beta^2+\bar{q}^2/4)}= \nonumber \\ &&
-\frac{1}{\beta^2}\ln{\left(1-e^{-2\beta c}\right)}-\frac{\bar{q}^2}{4\beta^2}\sum_{k=1}^\infty \frac{e^{-2k\beta c}}{k(k^2\beta^2+\bar{q}^2/4)}.
\label{B15}
\end{eqnarray}
The final expression of the Fourier image of the Firsov potential used in our programs is
\begin{equation}
\tilde U(\tilde{q})=4\pi Z e b^2a_F^2\beta^2A(\bar{q}),\;\; \bar{q}=bq=\frac{b}{a_B}\tilde{q},\;\; b=\frac{m_\mu}{2m_e}\left(\frac{9\pi^{2}}{128Z}\right)^{1/3}a_B.
\label{B16}
\end{equation}
Convergence in (\ref{B15}) is slow. Therefore, a sufficient number of terms must be included to obtain reliable results. In our programs, we have found that including one hundred terms is sufficient.

\section{Derivatives of Chebyshev polynomials at x=1}
\label{AppC}
From the definitions of Chebyshev polynomials of the first and second kind
\begin{equation}
T_n(x)=\cos{n\theta},\;\;U_n(x)=\frac{\sin{(n+1)\theta}}{\sin{\theta}},\;\;x=\cos{\theta},
\label{C1}
\end{equation}
we easily get
\begin{equation}
\frac{dT_n}{dx}=nU_{n-1},\;\;\frac{dU_n}{dx}=\frac{(n+1)T_{n+1}-xU_n}{x^2-1}.
\label{C2}
\end{equation}
To get the $x\to 1$ limit of the second equation, we apply L'H\^{o}pital's rule:
\begin{equation}
\left . \frac{dU_n}{dx}\right |_{x=1}=\frac{1}{2}\left[ (n+1)\frac{dT_{n+1}}{dx}-U_n(x)-x\frac{dU_n}{dx}\right ]_{x=1}.
\label{C3}
\end{equation}
Given the first equation in (\ref{C2}), it follows from (\ref{C3}) that
\begin{equation}
U^\prime_n(1)=\frac{1}{3}\left[ (n+1)^2-1\right]U_n(1)=\frac{1}{3}n(n+1)(n+2).
\label{C4}
\end{equation}
Another interesting, although somewhat more complex, way to obtain (\ref{C4}) is to use the formula for the $k$th derivative of Chebyshev polynomials \cite{Zhang_2004}
\begin{equation}
U^{(k)}_{n+k}(x)=2^kk!\sum\limits_{n_1+n_2+\ldots+n_{k+1}=n}U_{n_1}(x)U_{n_2}(x)\cdots U_{n_{k+1}}(x),
\label{C5}
\end{equation}
where the sum is over non-negative integers $n_1,n_2,\ldots,n_{k+1}$. In particular,
\begin{equation}
U^\prime_n(1)=2\sum_{k=1}^nU_{k-1}(1)U_{n-k}(1)=2\sum_{k=1}^n k(n+1-k),
\label{C6}
\end{equation}
and (\ref{C4}) follows from the well-known identities 
\begin{equation}
\sum_{k=1}^n k=\frac{1}{2}\,n(n+1) \;\;\mathrm{and}\;\; \sum_{k=1}^n k^2=\frac{1}{6}\,n(n+1)(2n+1).
\label{C7}
\end{equation}


\bibliographystyle{elsarticle-num}
\bibliography{foil.bib}

\end{document}